\documentclass[]{mn2e}

\usepackage[dvips]{graphicx,color}
\usepackage{times}

\usepackage{amsmath}
\usepackage{pifont}
\usepackage{lscape}

\def\nms{\mathsurround=0pt}
\def\oversim#1#2{\lower 2pt\vbox{\baselineskip 
0pt \lineskip 1pt
    \ialign{$\nms#1\hfil##\hfil$\crcr#2\crcr\sim\crcr}}}
\def\gtsim{\mathrel{\mathpalette\oversim>}} 
 \begin{document}

\title[Very massive stars in R136 and NGC 3603]
{The R136 star cluster hosts several stars whose individual masses greatly
exceed the accepted 150 M$_{\odot}$ stellar mass limit}
\author[P. A. Crowther et al.]{Paul A. 
Crowther$^{1}$\thanks{Paul.Crowther@shef.ac.uk},  
Olivier Schnurr$^{1,2}$, Raphael Hirschi$^{3, 4}$, Norhasliza Yusof$^{5}$, 
\newauthor Richard J. Parker$^{1}$, Simon P. Goodwin$^{1}$, Hasan Abu 
Kassim$^{5}$
\vspace{3mm} \\ 
$^{1}$Department of Physics and Astronomy, University of Sheffield, 
Sheffield S3 7RH, UK\\
$^{2}$ Astrophysikalisches Institut Potsdam, An der Sternwarte 16, D-14482 Potsdam, Germany\\
$^{3}$ Astrophysics Group, EPSAM, University of Keele, Lennard-Jones Labs, Keele, ST5 5BG, UK\\
$^{4}$ Institute for the Physics and Mathematics of the Universe, 
University of Tokyo, 5-1-5 Kashiwanoha, Kashiwa, 277-8583, Japan \\
$^{5}$ Department of Physics, University of Malaya, 50603 Kuala Kumpur, 
Malaysia }
\date{\today}

\pagerange{\pageref{firstpage}--\pageref{lastpage}} \pubyear{2010}

\maketitle

\label{firstpage}

\begin{abstract} Spectroscopic analyses of hydrogen-rich WN5--6 stars 
within  the
young star clusters NGC 3603 and R136 
are presented, 
using archival Hubble Space  Telescope and Very Large Telescope 
spectroscopy, and high spatial 
resolution near-IR photometry, including Multi-Conjugate Adaptive Optics 
Demonstrator (MAD) imaging of ~R136.  We derive high stellar 
temperatures for the WN stars in NGC 3603 ($T_{\ast} \sim$ 42$\pm$2 kK) 
and R136 ($T_{\ast} \sim$ 53$\pm$ 3 kK) plus clumping-corrected mass-loss
rates of 2 -- 5 $\times 10^{-5}$ M$_{\odot}$\,yr$^{-1}$ which closely 
agree with theoretical predictions from Vink et al. These stars make a 
disproportionate  contribution to the global ionizing and mechanical wind 
power budget of their host clusters. Indeed,  R136a1 alone  supplies 
$\sim$7\% of the ionizing flux of the entire 30 Doradus region. 
Comparisons with  stellar  models calculated for the main-sequence 
evolution of 85 -- 500 
M$_{\odot}$ accounting for rotation suggest ages of $\sim$1.5 Myr and
initial masses in the range 105 -- 170 M$_{\odot}$ for three systems in NGC 
3603, plus 165  -- 320 M$_{\odot}$ for four stars in R136. Our high 
stellar masses are supported by consistent spectroscopic and 
dynamical mass determinations for the components of NGC 3603 A1. 
We consider the predicted X-ray luminosity of the R136 
stars if they were close, colliding wind binaries. 
R136c is consistent with a colliding wind binary system. However, short 
period, colliding wind systems are excluded for R136a WN stars if 
mass ratios are of order unity. Widely separated systems 
would have been expected to harden owing to early dynamical encounters 
with other  massive  stars within such a high density environment. From 
simulated star clusters, whose constituents are randomly sampled  from 
the Kroupa initial mass function, both NGC 3603 and  R136 are consistent 
with an tentative upper mass limit of $\sim$300  M$_{\odot}$. The 
Arches cluster is either too old to be used  to diagnose the upper mass 
limit, exhibits a deficiency of very massive  stars, or more likely
stellar masses have been underestimated -- initial masses for the 
most luminous stars in the Arches cluster approach 200  M$_{\odot}$ 
according to contemporary stellar and photometric results.  The 
potential for stars greatly exceeding 150 M$_{\odot}$ within 
metal-poor galaxies  suggests that such pair-instability supernovae could 
occur within the  local universe, as has been claimed for SN 2007bi. 
\end{abstract}

\begin{keywords}
binaries: general --  stars: early-type -- 
stars: fundamental parameters -- 
stars: Wolf-Rayet -- galaxies: star clusters: individual (R136, NGC 3603, Arches)
\end{keywords}

\section{Introduction}

Although the formation of very massive stars remains an unsolved problem 
of astrophysics (Zinnecker \& Yorke 2007), the past decade has seen a 
shift from a belief that there is no observational upper stellar mass 
cutoff (e.g. Massey 2003) to the widespread acceptance of a limit close 
to  150 M$_{\odot}$ (Figer 2005; Koen 2006). If stars above this limit 
were to exist, they would be exclusive to the youngest, highest mass 
star 
clusters, which are very compact (Figer 2005). However, spatially resolved 
imaging of such clusters is currently exclusive to the Milky Way and its 
satellite galaxies, where they are very rare. In addition, 
the accurate determination of stellar masses generally relies upon
spectroscopic and evolutionary models, unless the star is a member of 
an eclipsing 
binary system (Moffat 2008), something which occurs extremely rarely.

Most, if not all, stars form in groups or clusters (Lada \& Lada 2003). An 
average star forms with an initial mass of $\sim$0.5 M$_{\odot}$ while the 
relative proportion of stars of higher and lower mass obeys an apparently 
universal initial mass function (IMF, Kroupa 2002). In addition, there 
appears to be a relationship between the mass of a cluster and its 
highest-mass star (Weidner \& Kroupa 2006, Weidner et al. 2010). High mass 
stars ($>$8 M$_{\odot}$), ultimately leading to core-collapse supernovae, 
usually form in clusters exceeding ~100 M~$_{\odot}$ while stars 
approaching 150 M$_{\odot}$ have been detected in two ~10$^{4}$ 
M$_{\odot}$ Milky Way clusters, namely the Arches (Figer 2005; Martins et 
al. 2008) and NGC 3603 (Schnurr et al. 2008a). Is this 150 M$_{\odot}$ 
limit statistical or physical?

The determination of stellar masses of very massive stars from 
colour-magnitude diagrams is highly unreliable, plus corrections to 
present mass estimates need to be applied to estimate initial masses. 
Studies are further hindered by severe spatial crowding within the cores 
of these star clusters (Ma\'{i}z Apell\'{a}niz 2008). In addition, 
sophisticated techniques are required to extract the physical parameters 
of early-type stars possessing powerful stellar winds from 
optical/infrared spectroscopy (Conti et al. 2008). In general, the 
incorporation of line blanketing has led to a reduction in derived 
temperatures for O stars from photospheric lines (Puls et al. 2008) while 
the reverse is true for emission lines in Wolf-Rayet stars (Crowther 
2007).

Here we re-analyse the brightest members of the star cluster (HD 97950) 
responsible for NGC 3603, motivated in part by the case of A1 which is an 
eclipsing binary system, whose individual components have been measured by 
Schnurr et al. (2008a). We also re-analyse the brightest members of R136 
(HD 38268) - the central ionizing cluster of the Tarantula nebula (30 
Doradus) within the Large Magellanic Cloud (LMC). R136 is sufficiently 
young (Massey \& Hunter 1998) and massive ($\leq$5.5 $\times$ 10$^{4}$ 
M$_{\odot}$, Hunter et al. 1995) within the Local Group of galaxies to 
investigate the possibility of a physical limit beyond 150 M$_{\odot}$ 
(Massey \& Hunter 1998, Selman et al. 1999; Oey \& Clarke 2005).

As recently as the 1980s, component `a' within R136 was believed to be a 
single star with a mass of several thousand solar masses (Cassinelli et 
al. 1981; Savage et al. 1983), although others favoured a dense star 
cluster (Moffat \& Seggewiss 1983). The latter scenario was supported by 
speckle interferometric observations (Weigelt \& Bauer 1985) and 
subsequently confirmed by  Hubble Space Telescope (HST) imaging (Hunter 
et al. 1995). 

\begin{table}
\begin{center}
\caption{Log of spectroscopic observations of R136 and NGC~3603 stars used in this study}
\label{log}
\begin{tabular}{l@{\hspace{1.5mm}}l@{\hspace{1.5mm}}l@{\hspace{1.5mm}}l
@{\hspace{2mm}}l}
\hline
Star & Instrument & Grating & Date & Proposal/PI \\
\hline
R136a1 & HST/HRS  & G140L   & 1994 Jul & 5157/Ebbets  \\
R136a3 & HST/HRS  & G140L   & 1994 Jul & 5157/Ebbets  \\
\vspace{1mm}
R136a1 & HST/FOS  & G400H, & 1996 Jan & 6018/Heap  \\
       &          & G570H         &               &       \\
R136a2 & HST/FOS  & G400H, & 1996 Jan & 6018/Heap  \\
       &          & G570H         &                &       \\
R136a3 & HST/FOS  & G400H, & 1996 Jan & 6018/Heap  \\
       &          & G570H         &                &       \\
R136c  & HST/FOS  & G400H         & 1996 Nov & 6417/Massey  \\
\vspace{1mm}
R136a1+a2 & VLT/SINFONI &  K    & 2005 Nov-, & 076.D-0563/Schnurr  \\
          &             &       & 2005 Dec  \\
R136a3 & VLT/SINFONI &  K    & 2005 Nov- & 076.D-0563/Schnurr  \\
          &             &       & 2005 Dec  \\
R136c  & VLT/SINFONI &  K    & 2005 Nov- & 076.D-0563/Schnurr  \\
          &             &       & 2005 Dec  \\
\vspace{1mm}
NGC 3603A1 & HST/FOS  & G400H  & 1994 Sep  & 5445/Drissen \\
NGC 3603B & HST/FOS  & G400H  & 1994 Sep  & 5445/Drissen  \\
NGC 3603C & HST/FOS  & G400H  & 1994 Sep  & 5445/Drissen  \\
\vspace{1mm}
NGC 3603A1 & VLT/SINFONI  & K  & 2005 Apr- & 075.D-0577/Moffat  \\
          &             &       & 2006 Feb  \\
NGC 3603B & VLT/SINFONI  & K  & 2005 Apr-  & 075.D-0577/Moffat  \\
          &             &       & 2006 Feb  \\
NGC 3603C & VLT/SINFONI  & K  & 2005 Apr-  & 075.D-0577/Moffat  \\
          &             &       & 2006 Feb  \\
\hline
\end{tabular}
\end{center}
\end{table}

Here, we present new analyses of the brightest sources of NGC 3603 and
R136. In  contrast to classical Wolf-Rayet stars, these WN stars are 
believed to be  young, main-sequence stars, albeit possessing very strong 
stellar winds as a  result of their high stellar luminosities (Crowther 
2007). Section 2 describes archival HST
and Very Large Telescope (VLT) observations,
while the spectroscopic analysis is presented in Sect~3. Comparisons
with contemporary evolutionary models are presented in Sect. 4,
revealing spectroscopic masses in excellent agreement with dynamical
masses of the components of NGC 3603 A1. For the R136 stars, exceptionally
high initial masses of 165 -- 320 M$_{\odot}$ are inferred.
We consider the possibility  that these stars are binaries in Sect 5, 
from archival X-ray observations of R136. In Sect. 6 we simulate star 
clusters to re-evaluate the stellar  
upper mass limit, and find that both R136 and NGC 3603 are consistent with
a tentative upper limit of $\sim$ 300M$_{\odot}$. 
Sect.~7 considers the contribution of the R136 stars to the global properties of both 
this cluster and 30 Doradus. Finally, we consider the broader 
significance of an increased mass limit for stars in Sect.~8.

\begin{figure}
\begin{center}
\includegraphics[width=1.\columnwidth,clip]{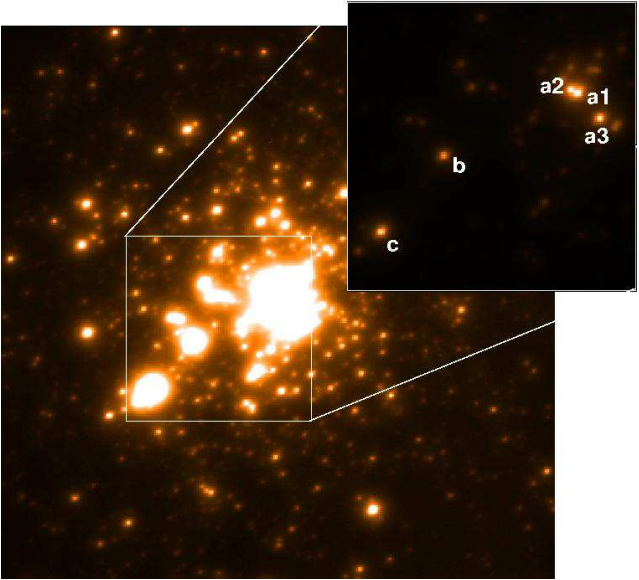}
\caption{
VLT MAD K$_{s}$-band 12 $\times$ 12 arcsec (3 $\times$ 3 parsec for the 
LMC distance of 49 kpc) image of R136 (Campbell et al. (2010) together with a 
view of the central 4 $\times$ 4 arcsec (1 $\times$ 1 parsec) in which the 
very massive WN5h stars discussed in this letter are labelled (component b 
is a lower mass WN9h star). Relative photometry agrees closely with 
integral field SINFONI observations (Schnurr et al. 2009).}
\label{MAD_K1}
\end{center}
\end{figure}

\begin{table}
\begin{center}
\caption{Stars brighter than M$_{\rm K_{s}} \sim$  --6 mag within 13 arcsec 
(0.5 parsec) of NGC 3603 A1, 
together with V-band photometry from Melena et al. (2008) in parenthesis.}
\label{ngc3603_photom}
\begin{tabular}{l@{\hspace{1.5mm}}l@{\hspace{1.5mm}}r@{\hspace{1.5mm}}r
@{\hspace{2mm}}r@{\hspace{2mm}}c}
\hline 
Name &   Sp Type & m$_{K_{s}}$  &   A$_{K_{s}}$  &   M$_{K_{s}}^{a}$  &  Binary$^{b}$\\
     &           &(m$_{V}$)      &  (A$_{V}$)    & (M$_{V}$)  & \\
\hline
A1  &    WN6h  &  7.42 $\pm$ 0.05 & 0.59 $\pm$ 0.03 & --7.57 $\pm$ 0.12 &
Yes\\
    &          & (11.18)            & (4.91$\pm$0.25)            & 
(--8.13$\pm$0.27)  \\
B   &    WN6h  &  7.42 $\pm$ 0.05 & 0.56 $\pm$ 0.03 & --7.54 $\pm$ 0.12 &
No?\\
    &          & (11.33)            & (4.70$\pm$0.25)            & 
(--7.77$\pm$0.27)  \\
C   &    WN6h$^{c}$  &  8.28 $\pm$ 0.05 & 0.56 $\pm$ 0.03 & --6.68 $\pm$ 0.12 &
Yes\\
    &          & (11.89)            & (4.66$\pm$0.25)            & 
(--7.17$\pm$0.27)  \\
\hline
\end{tabular}
\end{center}
\begin{small}
(a) For a distance of 7.6 $\pm$ 0.35 kpc (distance modulus 14.4 $\pm$ 
0.1 mag)\\
(b) A1 is a 3.77 day double-eclipsing system, while C is a 8.9 day SB1 
(Schnurr et al. 2008a) \\
(c) An updated classification scheme for Of, Of/WN and WN stars (N.R. Walborn,
\& P.A. Crowther, in preparation) favours O3\,If*/WN6 for NGC 3603 C
\end{small}
\end{table}

\section{Observations}

Our analysis of WN stars in NGC 3603 and R136 
is based upon archival UV/optical HST and near-IR 
VLT spectroscopy, summarised in Table~\ref{log}, combined 
with archival high spatial resolution near-IR imaging. We prefer 
the latter to optical imaging due to reduced extinction, efficient
correction for severe spatial crowding (Schnurr et al. 2008a,
2009) and 
consistency with studies of other young, high mass clusters
(e.g. Arches, Martins et al. 2008).

\subsection{NGC 3603}

Our spectroscopic analysis of the three
WN6h systems within NGC 3603 is based upon
archival HST/FOS spectroscopy from Drissen et al. (1995) plus
integral field VLT/SINFONI near-IR spectroscopy from Schnurr
et al. (2008a) 
-- see Table~\ref{log} for the log of observations.
The spectral resolution of the SINFONI datasets 
is R$\sim$3000, with adaptive optics (AO) used to observe A1, B and C, 
versus R$\sim$1300 for FOS for which a circular aperture of diameter 
0.26 arcsec was used.

Differential photometry from VLT/SINFONI 
integral field  observations (Schnurr et al 2008a), were tied to 
unpublished VLT/ISAAC K$_{s}$-band 
acquisition images from 16 Jun 2002 (ESO Programme 69.D-0284(A), P.I. 
Crowther) using the relatively isolated star NGC 3603 C. The VLT/ISAAC 
frames are calibrated against 2MASS  photometry  (Skrutskie et al. 2006) 
using 6 stars in common ($\pm$0.05 mag). These are presented in 
Table~\ref{ngc3603_photom} together with absolute magnitudes resulting
from  a distance of 7.6$\pm$0.35 kpc (distance modulus 14.4$\pm$0.1 
mag) plus  an extinction law from Melena et al. (2008) using
\[ A_{\rm V} = 1.1 R_{\rm V}^{\rm MW} + (0.29 - 0.35) R_{\rm V}^{\rm 
NGC3603} 
\]
where R$_{\rm V}^{\rm MW}$ = 3.1 for the Milky Way foreground component 
and R$_{\rm V}^{\rm NGC3603}$ =  4.3 for the internal NGC 3603 component.
We obtain A$_{K_{s}}$ = 0.12 A$_{V} \sim$ 0.56  -- 0.59  mag from 
spectral energy distribution fits, which are consistent with recent 
determinations of E(B-V) = 1.39 mag from Melena et al. (2008). In 
their  analysis, Crowther \& Dessart (1998) used a lower overall 
extinction of E(B-V) = 1.23 mag, albeit a higher distance of 10 kpc 
(distance modulus of 15.0 mag) to NGC 3603.

\begin{figure*}
\begin{center}
\leavevmode
\includegraphics[width=1.5\columnwidth,clip]{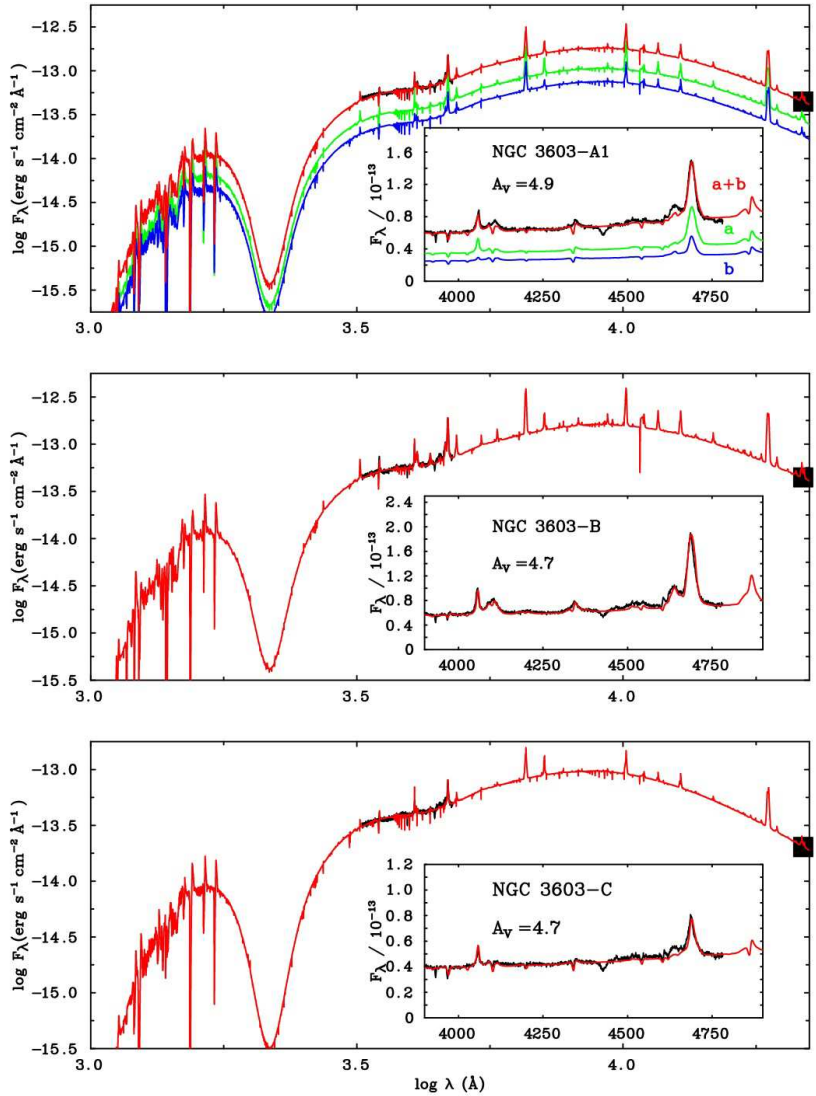}
\hfill \parbox[b]{4.75cm}{\caption{Spectral energy distributions of NGC 3603 WN6h stars from HST/FOS 
together with K$_{s}$ photometry from VLT/SINFONI and VLT/ISAAC plus reddened  
theoretical  spectral energy distributions (red lines).
Components  A1a (green lines) and A1b  (blue lines) are also shown 
separately.}
\label{ngc3603_sed}
}
\end{center}
\end{figure*}

\subsection{R136}

Our UV/optical/infrared spectroscopic analysis of all four
hydrogen-rich WN5h Wolf-Rayet (WR) stars in R136 (Crowther \& 
Dessart 1998) is also based upon archival HST and VLT spectroscopy
-- see Table~\ref{log} for the log of observations. 

Goddard High Resolution Spectrograph (GHRS) ultraviolet observations of
R136a stars have 
been described by de Koter et al. (1997), achieving a spectral resolution 
of R$\sim$2000--3000 using the SSA aperture of 0.22$\times$0.22 arcsec. This 
prevented individual UV spectroscopy for R136a1 and a2 which are separated 
by $\sim$0.1 arcsec. For context, 0.1 arcsec subtends 5000 AU at the 
distance of the LMC. R136a2 UV spectroscopy was attempted by S.~Heap (Programme
5297), but was assessed to be unreliable by de Koter et al. (1997) and so
is also excluded here. Note also that R136c has not been observed with 
GHRS. Visual Faint Object Spectroscopy (FOS) datasets, with a circular 
aperture of diameter 0.26 arcsec achieved R$\sim$1300, although R136a1 and 
R136a2 once again suffer significant contamination from one another. It is 
solely at near-IR wavelengths (SINFONI) that  R136a1 and a2 are 
spectrally separated,  for which  R136b served as an AO reference star 
(Schnurr et al. 2009).
The spectral resolution of the SINFONI datasets is R$\sim$3000. 

%
%

We employ high spatial resolution K$_{s}$-band photometry of 
R136. Differential K$_{s}$ photometry from AO assisted VLT/SINFONI integral
field datasets are tied to identical spatial resolution 
wider field VLT Multi-Conjugate 
Adaptive Optics Demonstrator (MAD) imaging (Campbell et al. 2010) using 
the relatively isolated star R136b (WN9h). Three overlapping Fields were 
observed with VLT/MAD, for which Field 1 provided the highest quality in 
R136 (FWHM $\sim$ 0.1 arcsec), 
as shown in Fig.~\ref{MAD_K1}, itself calibrated using archival 
HAWK-I and 2MASS datasets (see Campbell et al. 2010). For 12 stars in 
common between MAD photometry and HST Near Infrared Camera and Multi 
Object Spectrometer 
(NICMOS) F205W imaging (Brandner et al. 2001)
 transformed into the CTIO K-band system, Campbell et al. (2010) 
find  m$_{\rm K_{s}}$ (MAD) - m$_{\rm K}$ (HST) = -- 0.04 $\pm$0.05 mag. 

In Table~\ref{r136_photom} we present K$_{s}$ aperture photometry and 
inferred absolute magnitudes of stars brighter than $M_{\rm K_{s}} \sim$ 
--6 mag within 20 arcsec (5 parsec) of R136a1.  Of these only R136a, 
b and c components lie within a projected distance of 1 pc from 
R136a1. Spectral  types are taken from Crowther \& Dessart (1998) or 
Walborn \& Blades  (1997), although other authors prefer alternative 
nomenclature, e.g.  O2\,If for both Mk\,39 and R136a5 according to Massey 
et al. (2004, 2005).

Interstellar 
extinctions for the
R136 WN5h stars are derived from UV to near-IR spectral energy 
distribution fits 
(R136c lacks UV spectroscopy), adopting foreground Milky Way (LMC) extinctions of 
A$_{\rm K_{s}}$ = 0.025 (0.06) mag, plus variable internal 30 
Doradus 
nebular extinction. The adopted extinction law
follows  Fitzpatrick \& Savage (1984) as follows
\[ A_{\rm V} =  0.07 R_{\rm V}^{\rm MW}  + 0.16 R_{\rm V}^{\rm LMC}
+ (0.25-0.45) R_{\rm V}^{\rm 30Dor} \]
where R$_{\rm V}^{\rm MW}$ = R$_{\rm V}^{\rm LMC}$ = 3.2 and 
R$_{\rm V}^{\rm 30Dor}$ = 4.0.
We derive A$_{\rm K_{s}} 
\sim$0.22 mag for the R136a stars and 0.30 mag for R136c and adopt A$_{\rm 
K_{s}} =$ 0.20 mag for other stars except that Mk\,34 (WN5h) mirrors the 
higher extinction of  R136c.
From Table~\ref{r136_photom}, R136c is
0.5 mag fainter than R136a2 in the V-band (Hunter et al. 1995) but is 0.06
mag brighter in the K$_{s}$-band, justifying the higher extinction. An
analysis based  solely upon optical photometry could potentially 
underestimate the bolometric magnitude for R136c with respect to 
R136a2.
The main source of uncertainty in absolute 
magnitude results from  the distance to the LMC. The mean of 7 independent 
techniques (Gibson  2000) suggests an LMC  distance modulus of 18.45 $\pm$ 
0.06 mag. Here we  adopt an uncertainty of $\pm$0.18 mag owing to systematic 
inconsistencies between the various methods.

\begin{table}
\begin{center}
\caption{Stars brighter than M$_{K_{s}} \sim$  --6 mag within 20 arcsec 
(5 parsec) of R136a1, together with V-band photometry from Hunter et 
al. (1995) in parenthesis. 
}
\label{r136_photom}
\begin{tabular}{l@{\hspace{0.5mm}}l@{\hspace{1.5mm}}r@{\hspace{1.5mm}}r
@{\hspace{2mm}}r@{\hspace{2mm}}c}
\hline 
Name &   Sp Type & m$_{K_{s}}$  &   A$_{K_{s}}$  &   M$_{K_{s}}^{a}$  &  Binary$^{b}$\\
     &           &(m$_{V}$)      &  (A$_{V}$)    & (M$_{V}$)  & \\
\hline
R134    &WN6(h)  &10.91 $\pm$ 0.09 & 0.21 $\pm$ 0.04 & --7.75 $\pm$ 0.20& No?\\
        &        &(12.89$\pm$0.08)           & (1.77$\pm$0.33)           & 
(--7.33$\pm$0.38)\\
R136a1  &WN5h    &11.10 $\pm$ 0.08 & 0.22 $\pm$ 0.02 & --7.57 $\pm$ 0.20& No?\\
        &        &(12.84$\pm$0.05)           & (1.80$\pm$0.17)           & 
(--7.41$\pm$0.25) \\
R136c   &WN5h    &11.34 $\pm$ 0.08& 0.30 $\pm$ 0.02  &--7.41 $\pm$ 0.20& Yes?\\
        &        &(13.47$\pm$0.08)           & (2.48$\pm$0.17)           & 
(--7.46$\pm$0.26) 
\\
R136a2  &WN5h    &11.40 $\pm$ 0.08 &0.23 $\pm$ 0.02  &--7.28 $\pm$ 0.20& No?\\
        &        &(12.96$\pm$0.05)           & (1.92$\pm$0.17)           & 
(--7.41$\pm$0.25) 
\\
Mk\,34    &WN5h    &11.68 $\pm$ 0.08 & 0.27 $\pm$ 0.04  &--7.04 $\pm$ 0.20& 
Yes\\
        &        & (13.30$\pm$0.06)         & (2.22$\pm$0.33)            & 
(--7.37$\pm$0.38) \\
R136a3  &WN5h    &11.73 $\pm$ 0.08 &0.21 $\pm$ 0.02  &--6.93 $\pm$ 0.20 &No?\\
        &       & (13.01$\pm$0.04)          & (1.72$\pm$0.17)            & 
(--7.16$\pm$0.25) \\
R136b   &WN9ha   &11.88 $\pm$ 0.08 &0.21 $\pm$ 0.04  &--6.78 $\pm$ 0.20& No?\\
        &        &(13.32$\pm$0.04)          & (1.74$\pm$0.33)            & 
(--6.87$\pm$0.38) \\
Mk\,39    &O2--3\,If/WN$^{c}$ &       12.08 $\pm$ 0.08 &0.18 $\pm$ 0.04 & --6.55 $\pm$ 
0.20 &
Yes\\
        &        & (13.01$\pm$0.08)          & (1.46$\pm$0.33)   & 
(--6.90$\pm$0.38) \\
Mk\,42    &O2--3\,If/WN$^{c}$  &      12.19 $\pm$ 0.08 &0.17 $\pm$ 0.04 & --6.43 
$\pm$ 0.20 & No?\\
        &        & (12.84$\pm$0.05)     & (1.38$\pm$0.33) & 
(--6.99$\pm$0.38) \\
Mk\,37a   &O4\,If+$^{c}$   &12.39 $\pm$ 0.11 &0.21 $\pm$ 0.04 & --6.27 $\pm$ 0.21& 
No?\\
        &        & (13.57$\pm$0.05)     & (1.74$\pm$0.33)   & 
(--6.62$\pm$0.38) \\
Mk\,37Wa  &O4\,If+   &12.39 $\pm$ 0.11 &0.19 $\pm$ 0.04 & --6.25 $\pm$ 0.21 
&?\\
        &        & (13.49$\pm$0.05)     & (1.62$\pm$0.33)   & 
(--6.58$\pm$0.38) \\
R136a5  &O2--3If/WN$^{c}$  &      12.66 $\pm$ 0.08 &0.21 $\pm$ 0.04 & --6.00 $\pm$ 
0.20 & No?\\
         &                 &  (13.93$\pm$0.04)    & (1.74$\pm$0.33) & 
(--6.26$\pm$0.38) \\
\hline
\end{tabular}
\end{center}
\begin{small}
(a) For a distance modulus of 18.45 $\pm$ 0.18 mag (49 $\pm$ 4 kpc)\\
(b) R136c is variable and X-ray bright, Mk\,39 is a 92-day SB1 
binary (Massey et al. 2002; Schnurr et al. 2008b) and Mk\,34 is a binary
according to unpublished Gemini observations (O. Schnurr et al. in 
preparation)\\
(c) An updated classification scheme for Of, Of/WN and 
WN stars (N.R. Walborn,
 \& P.A. Crowther, in preparation) favours O2\,If$^{\ast}$ for Mk\,42
and R136a5, O3.5\,If for Mk\,37a and O2\,If/WN for Mk\,39
\end{small}
\end{table}

\section{Spectroscopic Analysis}

For our spectroscopic study we employ the non-LTE atmosphere code CMFGEN 
(Hillier \& Miller 1998) which solves the radiative transfer equation in 
the co-moving frame, under the additional constraints of statistical and 
radiative equilibrium. Since CMFGEN does not solve the momentum equation, 
a density or velocity structure is required. For the supersonic part, the 
velocity is parameterized with an exponent of $\beta$ = 0.8. This is 
connected to a hydrostatic density structure at depth, such that the 
velocity and velocity gradient match at this interface. The subsonic 
velocity structure is defined by a fully line-blanketed, plane-parallel 
TLUSTY model (Lanz \& Hubeny 2003)  whose gravity is closest to that
obtained from stellar masses derived using evolutionary models,
namely $\log g = 4.0$ for R136 stars and $\log g = 3.75$ for NGC 3603 stars.
CMFGEN incorporates line 
blanketing through a super-level approximation, in which atomic levels of 
similar energies are grouped into a single super-level which is used to 
compute the atmospheric structure.

Stellar temperatures, T$_{\ast}$, correspond to a Rosseland 
optical depth 10, which is typically 1,000 K to 2,000 K higher than 
effective temperatures T$_{2/3}$ relating to optical depths of 2/3 in such 
stars.

Our model atom include the following ions: H I, He I-II, C III-IV, N 
III-V, O III-VI, Ne IV-V, Si IV, P IV-V, S IV-V, Ar V-VII, Fe IV-VII, Ni 
V-VII, totalling 1,141 super-levels (29,032 lines). Other than H, He, CNO 
elements, we adopt solar abundances (Asplund et al. 2009) for NGC 3603,
which are supported by nebular studies (Esteban et al. 2005, Lebouteiller et al. 2008).
LMC nebular abundances (Russell \& Dopita 1990) are adopted for R136, with
other metals scaled to 0.4 Z$_{\odot}$, also supported by 
nebular studies of 30 Doradus (e.g. Peimbert 2003, 
Lebouteiller et al. 2008). We have assumed a depth-independent Doppler 
profile for all lines when solving for the atmospheric structure in the 
co-moving frame, while in the observer's frame, we have adopted a uniform 
turbulence of 50 km s$^{-1}$. Incoherent electron scattering and Stark 
broadening for hydrogen and helium lines are adopted. With regard to wind 
clumping (Hillier 1991), this is incorporated using a radially-dependent 
volume filling factor, $f$, with  $f_{\infty}$ = 0.1 at $v_{\infty}$, 
resulting in a reduction in mass-loss rate by a factor of $\sqrt{(1/f)}$ $\sim$ 3.

\begin{figure}
\begin{center}
\includegraphics[width=0.6\columnwidth,angle=270,clip]{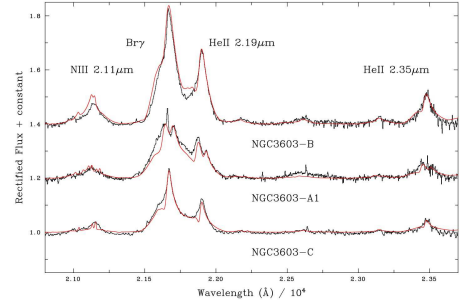}
\caption{Rectified, spatially resolved near-IR (VLT/SINFONI) spectroscopy 
of NGC 3603 WN6h stars (Schnurr et al. 2008a, black), including A1 close 
to  quadrature (A1a  blueshifted by  330 km s$^{-1}$ and A1b redshifted by 433 
km s$^{-1}$),  together  with synthetic spectra (red, broadened by 50 km\,s$^{-1}$). 
N\,{\sc iii}  2.103 -- 2.115$\mu$m and He\,{\sc i} 2.112 -- 
2.113$\mu$m contribute to the emission feature blueward of Br $\gamma$}
\label{ngc3603_k}
\end{center}
\end{figure}

\subsection{NGC 3603}

Fig.~\ref{ngc3603_sed} compares spectral energy distributions for the WN6h 
stars in NGC 3603 with reddened, theoretical models, including the 
individual components of A1. For this system we obtain M$_{\rm K_{s}}$ = 
--7.57 mag, such that we adopt $\Delta m$ = m$_{\rm A1a}$ -- m$_{\rm A1b}$ 
= --0.43 $\pm$0.3 mag for the individual components (Schnurr et al. 
2008a). We have estimated individual luminosities in two ways. First, on 
the basis of similar mean molecular weights $\mu$ for individual 
components, stellar luminosities result from L $\propto$ M$^{\alpha}$, 
with $\alpha \sim 1.5$ for ZAMS models in excess of 85 $M_{\odot}$. We 
derive L(A1a)/L(A1b) = 1.62$^{+0.6}_{-0.4}$ from the dynamical masses of 
116 $\pm$ 31 M$_{\odot}$ and 89 $\pm$ 16 M$_{\odot}$ obtained by Schnurr 
et al. (2008a) for A1a and A1b, respectively. Alternatively, we have 
obtained a luminosity ratio for components within A1 from a fit to the 
NICMOS lightcurve (Moffat et al. 2004, and E. Antokhina, priv. comm), 
using the radii of the Roche lobes and lightcurve derived temperatures. We 
also obtain L(A1a)/L(A1b) = 1.62 from this approach.

Although we have matched synthetic spectra to near-IR photometry, we note
that no major differences would be obtained with either HST Wide Field and
Planetary Camera 2 (WFPC2) or Advanced Camera for Surveys (ACS) datasets.
For example, our reddened spectral energy distribution for NGC 3603B implies
B = 12.40 mag, which matches HST/ACS photometry to within 0.06 mag 
(Melena et al. 2008). 


We have used diagnostic optical and near-IR lines, including N\,{\sc iii} 
4634-41, 2.103--2.115$\mu$m, N\,{\sc iv} 3478-83, 4058, together with 
He\,{\sc ii} 4686, 2.189$\mu$m plus Br$\gamma$. We are 
unable to employ helium temperature diagnostics since He\,{\sc i} lines 
are extremely weak at optical and near-IR wavelengths. Spectroscopic 
comparisons with HST/FOS datasets are presented in Fig.~\ref{ngc3603_sed}. 
Overall, emission features are well reproduced, although absorption 
components of higher Balmer-Pickering lines are too strong, especially
for NGC~3603C. Fits are similar to Crowther \& Dessart (1998), in spite of an
improved TLUSTY structure within the photosphere. Fortunately, our spectral
diagnostics are not especially sensitive to the details of the photosphere
since they assess the inner wind conditions. Spectroscopically, we 
use  VLT/SINFONI spectroscopy of A1a and A1b obtained close to quadrature 
(Schnurr 
et al. 2008a) to derive mass-loss rates and hydrogen contents. The 
ratio of He\,{\sc ii} 2.189$\mu$m to Br$\gamma$ provides an excellent 
diagnostic for the hydrogen content in WN stars, except for the latest 
subtypes (approximately WN8 and later). As such, we are able to use 
primarily optical spectroscopy for the determination of stellar 
temperatures, with hydrogen content obtained from near-IR spectroscopy.
Near-IR comparisons between synthetic spectra and observations are 
presented for each of the WN6h stars within NGC 3603 in 
Fig.~\ref{ngc3603_k}.


\begin{table}
\begin{center}
\caption{Physical Properties of NGC 3603 WN6h stars.}
\label{ngc3603_params}
\begin{tabular}{l@{\hspace{-3mm}}r@{\hspace{1.5mm}}r@{\hspace{1.5mm}}r
@{\hspace{2mm}}r}
\hline 
Name  & A1a   & A1b & B & C\\
\hline
T$_{\ast}$ (kK)$^{a}$ & 42 $\pm$ 2  & 40 $\pm$ 2 & 42 $\pm$ 2 & 44 $\pm$ 
2\\
$\log$ (L/L$_{\odot}$) & 6.39 $\pm$ 0.14 & 6.18 $\pm$ 0.14 & 6.46 $\pm$ 
0.07 
& 6.35 $\pm$ 0.07 \\
R$_{\tau = 2/3}$ (R$_{\odot}$) & 29.4$_{-4.3}^{+10.1}$ & 
25.9$_{-3.1}^{+7.2}$ & 33.8$_{-2.5}^{+2.7}$ & 26.2$_{-2.0}^{+2.1}$  \\
N$_{\rm LyC}$ (10$^{50}$ s$^{-1}$) & 1.6$_{-0.4}^{+0.8}$ & 
0.85$_{-0.23}^{+0.54}$ & 1.9$_{-0.3}^{+0.3}$ & 1.5$_{-0.3}^{+0.3}$ \\
$\dot{M}$ (10$^{-5}$ M$_{\odot}$ yr$^{-1}$) & 3.2$_{-0.6}^{+1.2}$ & 
1.9$_{-0.4}^{+0.9}$ &5.1$_{-0.6}^{+0.6}$ & 1.9$_{-0.2}^{+0.2}$ \\
$\log$ $\dot{M}$ - log $\dot{M}_{\rm Vink}^{c}$ & +0.14 & +0.24 & +0.22 & 
--0.04\\
V$_{\infty}$ (km s$^{-1}$) & 2600 $\pm$ 150 & 2600 $\pm$ 150 & 2300 $\pm$ 
150 & 2600 $\pm$ 150 \\
X$_{H}$ (\%) & 60 $\pm$ 5 & 70 $\pm$ 5 & 60 $\pm$ 5 & 70 $\pm$ 5 \\
M$_{\rm init}$ (M$_{\odot}$)$^{b}$ & 148$_{-27}^{+40}$ & 
106$_{-20}^{+23}$ & 166$_{-20}^{+20}$ & 137$_{-14}^{+17}$ \\
M$_{\rm current}$ (M$_{\odot}$)$^{b}$ & 120$_{-17}^{+26}$ & 
92$_{-15}^{+16}$ 
& 132$_{-13}^{+13}$ & 113$_{-8}^{+11}$ \\
M$_{K_{s}}$ (mag)$^{d}$ & --7.0 $\pm$ 0.3 & --6.6 $\pm$ 0.3 & --7.5 $\pm$ 0.1 & --6.7 $\pm$ 0.1 \\

\hline
\end{tabular}
\end{center}
\begin{small}
(a) Corresponds to the radius at a Rosseland optical depth of $\tau_{\rm 
Ross}$ = 10\\ (b) Component C is a 8.9 day period SB1 system (Schnurr et 
al. 2008a) \\ (c) dM/dt$_{\rm Vink}$ relates to Vink et al. (2001) 
mass-loss rates for Z = Z$_{\odot}$\\
(d) M$_{\rm K_{s}}$ = --7.57 $\pm$ 0.12 mag for A1, for which we adopt 
$\Delta$m = m$_{\rm A1a}$ - m$_{\rm A1b}$ = --0.43 $\pm$ 0.30 mag 
(Schnurr et al. 2008a). The ratio of their luminosities follows from
their dynamical mass ratios together with $L \propto \mu M^{1.5}$ 
(and is supported by NICMOS photometry from Moffat et al. 2004).
\end{small}
\end{table}

\begin{figure*}
\begin{center}
\leavevmode
\includegraphics[width=1.5\columnwidth,clip]{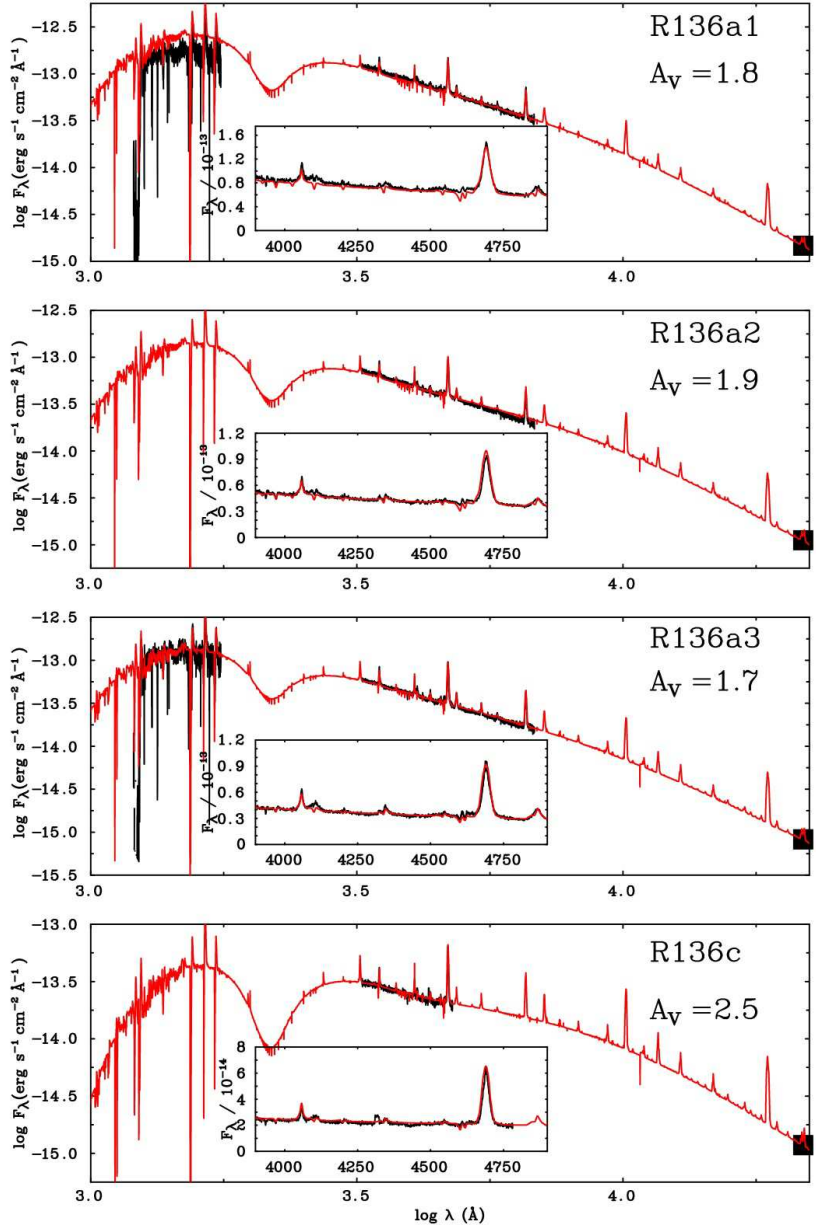}
\hfill \parbox[b]{4.75cm}{\caption{Spectral energy distributions of R136 WN5h stars from HST/FOS  
together using K$_{s}$ photometry from VLT/SINFONI calibrated with VLT/MAD 
imaging.  Reddened  theoretical spectral energy distributions 
are shown as red lines.}\label{r136_sed}}
\end{center}
\end{figure*}

Physical properties for these stars in NGC 3603 are 
presented in Table~\ref{ngc3603_params}, including evolutionary masses 
obtained from solar-metallicity non-rotating models. 
Errors
quoted in the Table account for both photometric and spectroscopic 
uncertainties, and represent the 
range of permitted values. However, for NGC 3603 C we assume that the 
primary dominates the systemic light since the companion is not
detected in SINFONI spectroscopy (Schnurr et al 2008a). Should the
companion make a non-negligible contribution to the integrated light, the 
derived properties set out in Table~\ref{ngc3603_params} would need to be 
corrected accordingly. The improved allowance for 
metal line blanketing  implies $\sim$10--20\% higher stellar temperatures 
(T$_{\ast}$ $\sim$ 40,000 -- 44,000 K) 
and, in turn, larger bolometric corrections (M$_{\rm Bol}$ -- M$_{\rm V}$ 
$\sim$ --3.7 $\pm$ 0.2 mag) for these stars than earlier studies (Crowther 
\& Dessart 1998), who adopted a different combination of E(B-V) and distance
for NGC 3603. We note that Schmutz \& Drissen (1999) have previously
derived T$_{\ast}$ $\sim$ 46,000 K for NGC 3603 B, resulting in a similar luminosity 
($\log L/L_{\odot}$ = 6.4) to that obtained here. Analysis of the NICMOS
lightcurve (Moffat et al. 2004) yields T(A1a)/T(A1b) = 1.06, in support 
of the spectroscopically-derived 
temperature ratio, albeit 7\% higher in absolute terms. In view of the underlying assumptions
of the photometric and spectroscopic techniques, overall consistency is satisfactory.

For A1a and B, we 
adopt abundances of X$_{\rm C}$ = 0.008\%, X$_{\rm N}$ = 0.8\% and X$_{\rm 
O}$ = 0.013\% by mass, as 
predicted by evolutionary models for X$_{\rm H}$ $\sim$ 60\%, versus 
X$_{\rm C}$ = 
0.005\%, X$_{\rm N}$ = 0.6\% and X$_{\rm O}$ = 0.25\% for A1b and C, for 
which X$_{\rm H}$ $\sim$ 70\%.

\subsection{R136}

Theoretical spectral energy distributions of all four WN5h stars
are presented in Fig.~\ref{r136_sed}. 
Although we have matched synthetic spectra to VLT/MAD + SINFONI
photometry, we note  that no significant differences would be obtained 
from WFPC2 imaging (e.g. 
Hunter et al. 1995), providing 
these are sufficiently isolated. For example, our reddened spectral 
energy distribution for R136a3 implies V = 
13.0 mag, in agreement with F555W photometry. VLT/MAD photometry is 
preferred to WFPC2 for the very crowded core of  R136 (e.g. a1 and a2), 
although recent Wide Field Camera 3 (WFC3) imaging of R136 achieves a 
similar spatial resolution at visible wavelengths. Nevertheless, we 
adhere to VLT/MAD + SINFONI for our primary photometric reference since 
\begin{enumerate}
\item Solely VLT/SINFONI spectrally resolves R136a1 from R136a2 
(Schnurr et al. 2009). We also note the consistent line to continuum
ratios between optical and near-IR diagnostics, ruling out potential 
late-type contaminants for the latter;
\item Uncertainties in dust 
extinction at K$_{s}$ are typically 0.02 mag versus $\sim$0.2 mag in the 
V-band, recalling $A_{K_{s}} \sim 0.12 A_{V}$.
R136c is significantly fainter than R136a2 at optical wavelengths
yet is brighter at K$_{s}$ and possesses a higher bolometric magnitude
(contrast these results with the optical study of R\"{u}hling 2008);
\item We seek to follow a consistent approach to recent studies of 
the Arches cluster which necessarily focused upon the K$_{s}$-band for
photometry and spectroscopy (e.g. Martins et al. 2008).
\end{enumerate}

\begin{table}
\begin{center}
\caption{Physical Properties of R136 WN5h stars.}
\label{r136_summary}
\begin{tabular}{l@{\hspace{-3mm}}r@{\hspace{1.5mm}}r@{\hspace{1.5mm}}r
@{\hspace{2mm}}r}
\hline 
Name  & a1   & a2 & a3 & c\\
BAT99 & 108  & 109 & 106 & 112 \\
\hline
T$_{\ast}$ (kK)$^{a}$ & 53 $\pm$3 & 53 $\pm$ 3 & 53 $\pm$ 3 & 51 $\pm$ 5\\
$\log$ (L/L$_{\odot}$) & 6.94 $\pm$ 0.09 & 6.78 $\pm$ 0.09 & 6.58 $\pm$ 0.09 & 6.75 $\pm$ 0.11 \\
R$_{\tau = 2/3}$ (R$_{\odot}$) & 35.4$_{-3.6}^{+4.0}$ & 
29.5$_{-3.0}^{+3.3}$ & 23.4$_{-2.4}^{+2.7}$ & 30.6$_{-3.7}^{+4.2}$  \\
N$_{\rm LyC}$ (10$^{50}$ s$^{-1}$) & 6.6$_{-1.3}^{+1.6}$ & 4.8$_{-0.7}^{+0.8}$ & 3.0$_{-0.4}^{+0.5}$ & 4.2$_{-0.6}^{+0.7}$ \\
$\dot{M}$ (10$^{-5}$ M$_{\odot}$ yr$^{-1}$) & 5.1$_{-0.8}^{+0.9}$ & 4.6$_{-0.7}^{+0.8}$ & 3.7$_{-0.5}^{+0.7}$ & 4.5$_{-0.8}^{+1.0}$ \\
$\log \dot{M}$ - log $\dot{M}_{\rm Vink}^{c}$ & +0.09 & +0.12 & +0.18 & +0.06\\
V$_{\infty}$ (km s$^{-1}$) & 2600 $\pm$ 150 & 2450 $\pm$ 150 & 2200 $\pm$ 150 & 1950 $\pm$ 150 \\
X$_{H}$ (\%) & 40 $\pm$ 5 & 35 $\pm$ 5 & 40 $\pm$ 5 & 30 $\pm$ 5 \\
M$_{\rm init}$ (M$_{\odot}$)$^{b}$ & 320$_{-40}^{+100}$ & 240$_{-45}^{+45}$ & 165$_{-30}^{+30}$ & 220$_{-45}^{+55}$ \\
M$_{\rm current}$ (M$_{\odot}$)$^{b}$ & 265$_{-35}^{+80}$ & 195$_{-35}^{+35}$ & 135$_{-20}^{+25}$ & 175$_{-35}^{+40}$ \\
M$_{K_{s}}$ (mag) & --7.6 $\pm$ 0.2 & --7.3 $\pm$ 0.2 & --6.9 $\pm$ 0.2 & --7.4 $\pm$ 0.2 \\
\hline
\end{tabular}
\end{center}
\begin{small}
(a) Corresponds to the radius at a Rosseland optical depth of $\tau_{\rm 
Ross}$ = 10\\ (b) Component R136c is probably a colliding-wind massive 
binary. For a mass ratio of unity, initial (current) masses of each 
component would correspond to $\sim$160 M$_{\odot}$ ($\sim$130 
M$_{\odot}$)\\ (c) dM/dt$_{\rm Vink}$ relates to Vink et al. (2001) 
mass-loss rates for Z = 0.43 Z$_{\odot}$
\end{small}
\end{table}

\begin{figure}
\begin{center}
\includegraphics[width=1.0\columnwidth,clip]{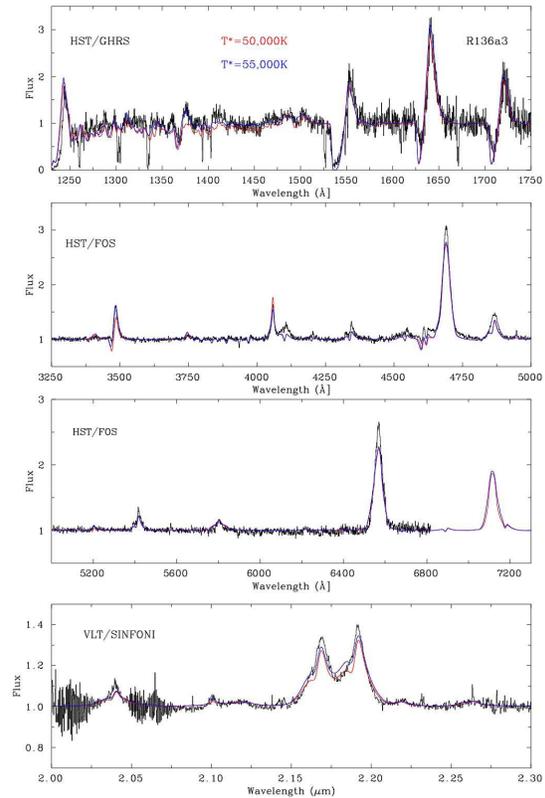}
\caption{Rectified, ultraviolet (HST/GHRS), visual (HST/FOS) and near-IR 
(VLT/SINFONI) spectroscopy of the WN5h star R136a3 together with 
synthetic UV, optical and near-infrared spectra, 
for T$_{\ast}$ = 50,000 K (red) and T$_{\ast}$ =55,000 K (blue). 
Instrumental broadening is accounted for, plus an additional
rotational broadening of 200 km\,s$^{-1}$.}
\label{r136a3_plot}
\end{center}
\end{figure}

We estimate terminal wind velocities from optical and near-IR 
helium lines. Ultraviolet HST/GHRS spectroscopy has been obtained for the 
R136a stars (Heap et al. 1994; de Koter et al. 1997), although the 
severe crowding within 
this region results in the spectra representing blends of individual 
components for R136a1 and a2 (apparent in Fig.~\ref{r136_sed}). 
As a result, we favour velocities 
from FOS and SINFONI spectroscopy for these cases.  

In Fig.~\ref{r136a3_plot} we present UV/optical/near-IR spectral fits for 
the WN5h star R136a3, which is sufficiently isolated that 
contamination from other members of R136a is minimal in each spectral region
(in contrast to R136a1 and a2 at UV/optical wavelengths).  Consistent
optical and near-IR fits demonstrate that no underlying red
sources contribute significantly to the K-band SINFONI datasets. 
Diagnostics
include O\,{\sc v} 1371, S\,{\sc v} 1501, N\,{\sc 
iv} 3478-83, 
4058, N\,{\sc v} 4603-20, 2.10$\mu$m together with He\,{\sc ii} 4686, 
2.189$\mu$m plus H$\alpha$, H$\beta$, Br$\gamma$. For T$_{\ast}$ $\leq$ 
50,000 K both O\,{\sc v} 1371 and N\,{\sc v} 2.11$\mu$m are underestimated 
for the R136a WN5h stars, while for T$_{\ast}$ $\geq$ 56,000 K, N\,{\sc 
iv} 3478-83 and 4057 become too weak, and S\,{\sc v} 1501 becomes too 
strong. Therefore, we favour T$_{\ast} \sim$53,000 $\pm$ 3,000 K, except 
that T$_{\ast} \sim$ 51,000 $\pm$5,000 K is preferred for R136c since UV 
spectroscopy is not available and N\,{\sc v} 2.100$\mu$m is weak/absent. 
Again, we are  unable to employ helium temperature diagnostics 
since He\,{\sc i} lines are extremely weak.

\begin{figure}
\begin{center}
\includegraphics[width=0.6\columnwidth,angle=270,clip]{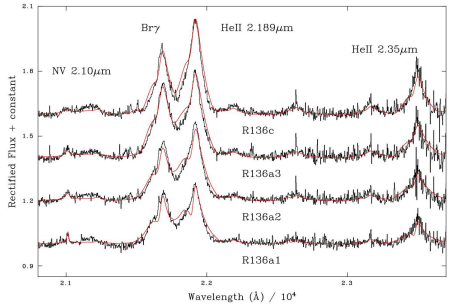}
\caption{Rectified, spatially resolved near-IR (VLT/SINFONI, Schnurr et 
al. 2009) spectroscopy  of R136 WN5h stars (black) together with synthetic 
spectra, accounting for instrumental broadening (100 km\,s$^{-1}$)
plus rotational broadening of 200 km\,s$^{-1}$ for R136a2, a3 and c. 
Consistent hydrogen contents are obtained from the peak 
intensity ratio of Br$\gamma$/HeII 2.189$\mu$m and optical (Pickering-Balmer series) 
diagnostics. Nitrogen emission includes N\,{\sc v} 2.100$\mu$m and N\,{\sc iii} 2.103 - 
2.115$\mu$m.}
\label{r136}
\end{center}
\end{figure}

Fig.~\ref{r136_sed} compares synthetic spectra with 
HST/FOS spectroscopy, for which emission features are again well matched, 
albeit with predicted Balmer-Pickering absorption components that are too 
strong. Fig.~\ref{r136} 
presents near infra-red spatially-resolved spectroscopy of R136 WN stars
(Schnurr et al.  2009) together with synthetic spectra, allowing for 
instrumental broadening (100 km\,s$^{-1}$) plus additional rotational 
broadening of 200 km\,s$^{-1}$ for R136a2, a3, c.
As for NGC 3603 stars, the ratio of He\,{\sc ii} 2.189$\mu$m to Br$\gamma$ provides
an excellent diagnostic for the hydrogen content for the R136 stars. A 
summary of the resulting physical and chemical parameters is presented in 
Table~\ref{r136_summary},  with errors once again accounting for both 
photometric and spectroscopic uncertainties, representing the 
range of permitted values, although the primary 
uncertainty involves the distance to the LMC.

With respect to earlier studies (Heap et al. 1994; de Koter et al. 1997; 
Crowther \& Dessart 1998), the improved allowance for metal line 
blanketing also infers 20\% higher stellar temperatures 
($T_{\ast} \sim$53,000 K) 
for these stars, and in turn, larger bolometric corrections (M$_{\rm Bol} 
- M_{\rm V}$ $\sim$ --4.6 mag).  Such differences, with respect to 
non-blanketed analyses, are typical of Wolf-Rayet stars (Crowther 2007). 
 Similar temperatures ($T_{\ast}$ = 50,000 or 56,000 K) 
 to the  present study were  obtained  for these 
WN stars  by R\"{u}hling (2008)  using a grid  calculated from the Potsdam 
line-blanketed atmospheric code (for a summary see Ruehling et al. 
2008), albeit with 0.1--0.2 dex lower luminosities  ($\log L/L_{\odot}$ = 
6.4 -- 6.7) using solely optically-derived (lower) extinctions and 
absolute magnitudes.

To illustrate the sensitivity of mass upon luminosity, we use the example 
of R136a5 (O2--3\,If/WN) for which we estimate $T_{\ast} \sim$ 50,000 K 
using the same diagnostics as the WN5h stars, in good agreement with 
recent analyses of O2 stars by Walborn et al. (2004) and Evans et al. 
(2010b). This reveals a luminosity of $\log L/L_{\odot}$ = 6.35, 
corresponding to an initial mass in excess of 100 M$_{\odot}$, versus 
$T_{\ast} \sim$ 42,500 K, $\log L/L_{\odot}$ = 5.9 and $\sim$65 
M$_{\odot}$ according to de Koter et al. (1997).

Regarding elemental abundances, we use the H$\beta$ to He\,{\sc ii}
5412 and Br$\gamma$ to He\,{\sc ii} 2.189$\mu$m ratios to derive (consistent) 
hydrogen contents, with the latter serving as the primary diagnostic for
consistency with NGC 3603 stars (current observations do not include
He\,{\sc ii} 5412).  We  adopt scaled solar abundances for all metals other than CNO elements. Nitrogen abundances 
of X$_{N}$ = 0.35\% by mass, as predicted by evolutionary models, are consistent with near-IR N\,{\sc v} 
2.100$\mu$m recombination line observations, while we adopt carbon and 
oxygen 
abundances 
of X$_{C}$ = X$_{O}$ = 0.004\% by mass.

\begin{figure*}
\begin{center}
\leavevmode
\includegraphics[width=1.5\columnwidth,clip]{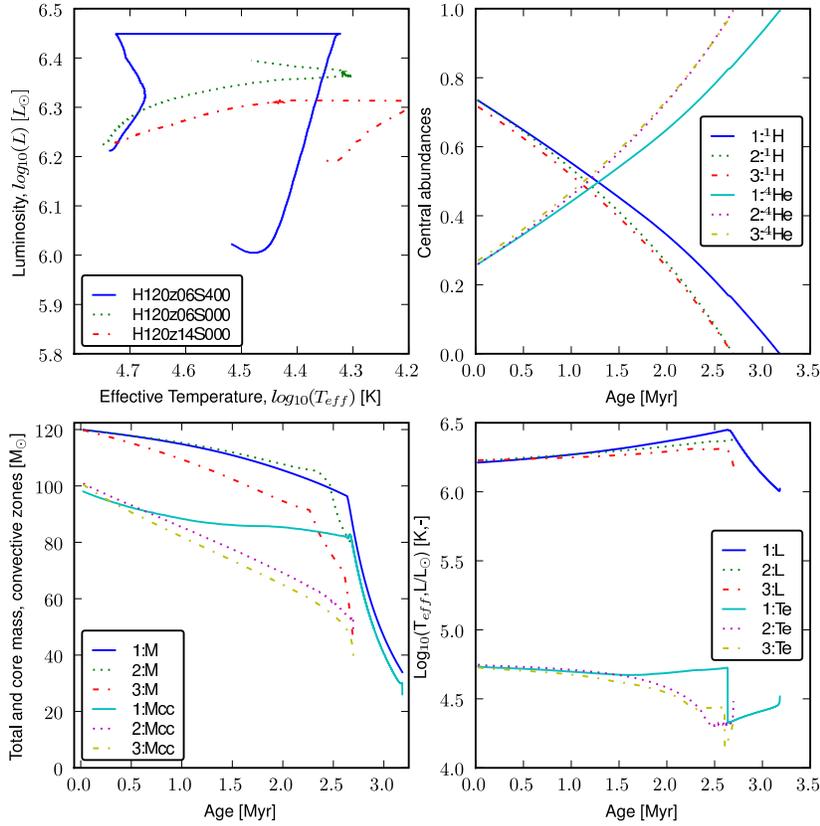}
\hfill \parbox[b]{4.75cm}{
\caption{Comparison between main sequence evolutionary predictions for both rotating (Z 
= 0.006: H120z06S400) and non-rotating (Z = 0.006: H120z06S000, Z = 0.014: H120z14S000) 
120 M$_{\odot}$ models. 
Horizontal lines in the top left-hand panel correspond to the transition to the Wolf-
Rayet phase.}\label{evol}
}
\end{center}
\end{figure*}

\section{Evolutionary Models}

We have calculated a grid of main-sequence models using the latest 
version of the Geneva  stellar evolution code for 85, 120, 150, 200, 300
and 500 M$_{\odot}$. The main-sequence evolution of such high-mass 
stars does not suffer from stability issues. 
Although a detailed description is provided 
elsewhere (Hirschi et al. 2004), together with recent updates (Eggenberger 
et al. 2008), models include the physics of rotation and mass loss, which 
are both crucial to model the evolution of very massive stars. 
Although many details relating to the evolution of very massive stars 
remain uncertain, we solely consider the main-sequence evolution here
using standard theoretical mass-loss prescription for O stars (Vink et 
al. 2001) for which Mokiem et al. (2007) provide supporting empirical
evidence. In the models, we consider the onset of the Wolf-Rayet phase to 
take place when the surface hydrogen content X$_{\rm H} <$ 30\% if 
T$_{\rm eff}$ $\geq$ 
10,000 K, 
during which an empirical mass-loss calibration is followed (Nugis \& 
Lamers 2000). The post-main sequence evolution is beyond the scope of this 
study and will be discussed elsewhere. 

All the main effects of rotation 
are included in the calculations: centrifugal support, mass-loss 
enhancement and especially mixing in radiative zones (Maeder 2009),
although predictions for both non-rotating and rotating models are
considered here. For rotating models we choose an initial ratio 
of the velocity to critical (maximum) rotation of $v_{\rm init}/v_{\rm 
crit}$ = 0.4, which corresponds 
to surface equatorial velocities of around 350 km s$^{-1}$ for the 85 
M$_{\odot}$ model and around 450 km s$^{-1}$ for the 500 M$_{\odot}$ case. 

We have calculated both solar (Z=1.4\% by mass) and LMC (Z=0.6\% by mass) 
metallicities. The evolution of 120 M$_{\odot}$ models in the 
Hertzsprung-Russell diagram is presented in Fig~\ref{evol}. The effects of 
rotation are significant.  Due to additional mixing, helium is mixed out 
of the core and thus the opacity in the outer layers decreases. This 
allows rotating stars to stay much hotter than non-rotating stars. Indeed, 
the effective temperature of the rotating models stay as high as 45,000 - 
55,000 K, whereas the effective temperature of non-rotating models 
decreases to 20,000 - 25,000 K. Therefore, rapidly rotating stars progress 
directly to the classical Wolf-Rayet phase, while slow rotators are 
expected to become $\eta$ Car-like Luminous Blue Variables (see Meynet \& 
Maeder 2005).

Rotating models can reach higher luminosities, especially at very
low  metallicity (see Langer et al. 2007). This is explained by additional 
mixing above the convective core. Finally, by comparing models at solar 
metallicity and LMC metallicity, we can see that lower metallicity models 
reach higher luminosities. This is due to weaker mass-loss at lower 
metallicity.

Langer et al. (2007) provide 150 M$_{\odot}$ tracks at 0.2 and 0.05 
Z$_{\odot}$ with high initial rotation velocities of 500 km\,s$^{-1}$. 
From these it is apparent that rapidly rotating very metal-deficient 
models can achieve high stellar luminosities. According to Langer et al. 
(2007) a 150 M$_{\odot}$ model reaches $\log L/L_{\odot} \sim$ 6.75 at 
0.05 Z$_{\odot}$ and 6.5 at 0.2 Z$_{\odot}$. Could the R136 stars 
represent rapidly rotating stars of initial mass 150 M$_{\odot}$? We have 
calculated a SMC metallicity (0.14 Z$_{\odot}$) model for a 150 
M$_{\odot}$ star initially rotating at $v_{\rm init}$ = 450 km\,s$^{-1}$, 
which achieves $\log L/L_{\odot} \approx$ 6.5 after 1.5 Myr and 6.67 after 
2.5 Myr. Our models are thus compatible with Langer et al. (2007) described above.
We nevertheless exclude the possibility that the R136 WN5 stars possess 
initial masses below 150 M$_{\odot}$ since:
\begin{enumerate}
\item the metallicity of 30 Doradus is a factor of three 
higher than the SMC (e.g. Peimbert 2003, Lebouteiller et al. 2008); 
\item R136 has an age of less than 2 Myr (de Koter et al. 1998, Massey \& 
Hunter 1998), since its massive stellar population is analogous to the 
young star clusters in Car OB1 (1--2 Myr, Walborn 2010) and NGC 3603
(this study); 
\item Clumping-corrected mass-loss rates for the R136 WN stars
agree well with LMC metallicity predictions (Vink et al. 2001), which
are also supported by studies of O stars in the Milky Way, LMC and SMC 
(Mokiem et al. 2007).
\end{enumerate}

\begin{figure*}
\begin{center}
\includegraphics[width=1\columnwidth,angle=270,clip]{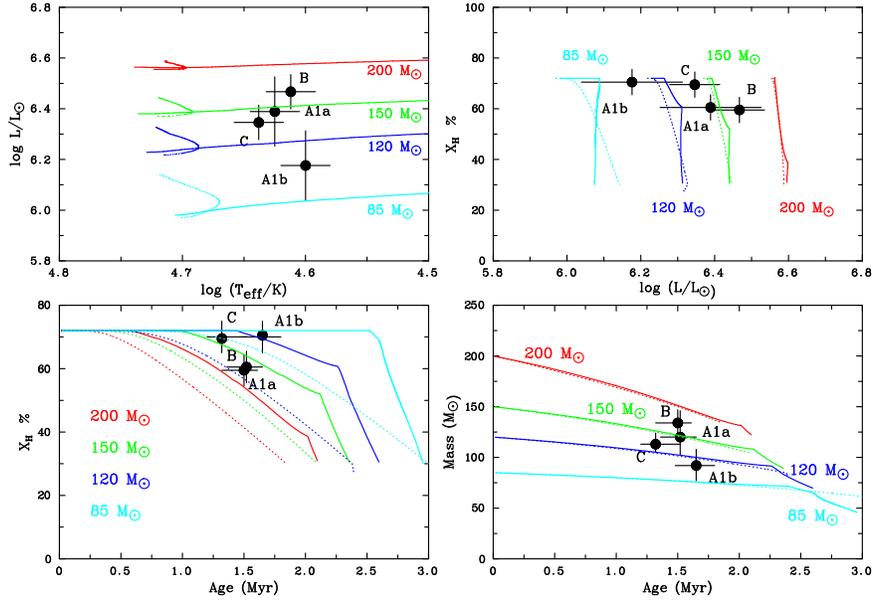}
\caption{Comparison between Solar metallicity (Z = 1.4\%) models 
calculated for the main-sequence evolution of 85 - 200 M$_{\odot}$ stars 
(initially rotating at $V_{\rm init}/v_{\rm crit}$ =0.4 [dotted] and 0 
[solid]), and the 
physical properties  derived from  spectroscopic analysis of NGC 3603 WN6h 
stars. We obtain excellent agreement with  dynamical masses for A1a and 
A1b for initially non-rotating models at ages of ~1.5 $\pm$ 0.1  Myr. 
Current mass-loss rates match solar-metallicity theoretical  predictions
(Vink et al. 2001) to within ~0.2 dex.}
\label{ngc3603_evol}
\end{center}
\end{figure*}

\begin{figure*}
\begin{center}
\includegraphics[width=1\columnwidth,angle=270,clip]{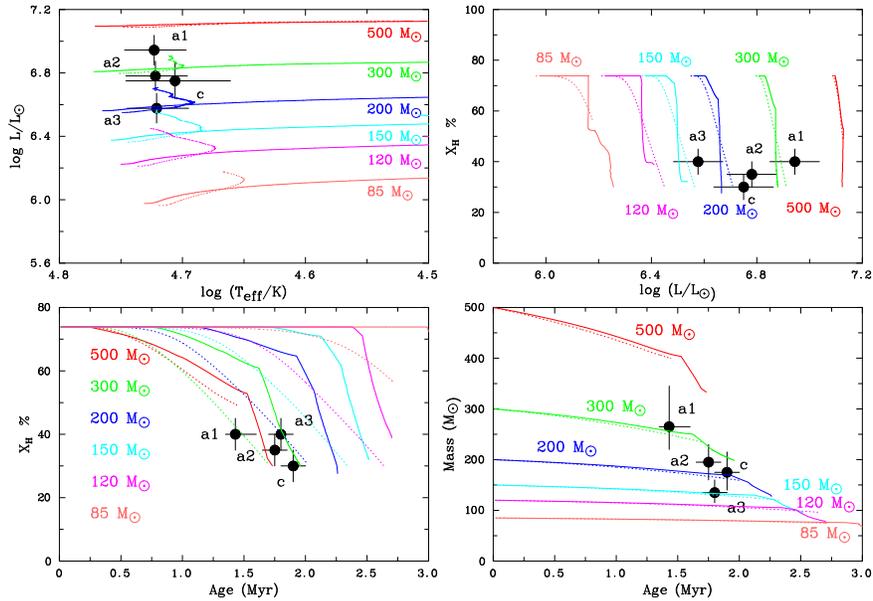}
\caption{Comparison between LMC-metallicity models calculated for the 
main-sequence evolution of 85 - 500 M$_{\odot}$ stars, initially rotating 
at $v_{\rm init}/v_{\rm crit}$ = 0.4 (dotted) or 0 (solid) and the physical 
properties derived from our spectroscopic analysis. We obtain excellent 
agreement for initially rapidly rotating, 165 -- 320 M$_{\odot}$ stars at 
ages of $\sim$1.7 $\pm$ 0.2 Myr. Current mass-loss rates match 
LMC-metallicity theoretical predictions (Vink et al. 2001) to within 0.2 
dex.}
\label{r136_fig1to4}
\end{center}
\end{figure*}

The surface abundances of a majority of high mass stars can be well 
reproduced by models of single stars including the effects of rotation. 
However, the VLT FLAMES survey has highlighted some discrepancies between 
models and observations (Hunter et al. 2008), questioning the efficiency 
of rotation induced mixing. We are currently investigating this matter by 
comparing the composition of light elements like boron and nitrogen 
between models and observations and we find that models including rotation 
induced mixing reproduces the abundances of most stars well (Frischknecht 
et al. 2010).

We can nevertheless consider how a less 
efficient rotation-induced mixing (or absence of mixing) would affect the 
conclusions of this paper. If rotation induced mixing was less efficient 
than our models predict, the masses derived here would remain at 
least as high, and would usually be higher. Indeed, less efficient mixing 
prevents the luminosity to increase as much since less helium is mixed up 
to the surface and the mean molecular weight, $\mu$, remains lower ($L 
\approx \mu M^{1.5}$). In particular, less efficient mixing would prevent 
stars with initial masses around or below 150 M$_{\odot}$ from reaching 
such high  luminosities as predicted in Langer et al. (2007, see 
discussion in  previous paragraph) and would bring further support to our 
claim that  these stars are more massive than 150 M$_{\odot}$. The 
challenge would  then be to explain the high effective temperature 
(T$_{\rm eff} \sim$ 50 kK) observed for the R136 stars, which is best 
explained by rotation  induced mixing. Note that less efficient mixing 
would also make impossible  the quasi-chemical evolution of fast rotating 
(single or binary) stars  (Yoon et al. 2006) which is currently one of the 
best scenarios for long/soft Gamma Ray Bursts progenitors.

\subsection{NGC 3603}

Fig.~\ref{ngc3603_evol} compares various observational properties of
NGC 3603 WN stars with solar metallicity evolutionary predictions.
Non-rotating models imply current masses of 120$_{-17}^{+26}$ M$_{\odot}$ 
and 92$_{-15}^{+16}$ M$_{\odot}$ for A1a and A1b, respectively, at an age 
of $\sim$1.5 $\pm$ 0.1 Myr. These are in excellent agreement with 
dynamical mass determinations of 116 $\pm$ 31 M$_{\odot}$ and 89 $\pm$ 16 
M$_{\odot}$ for the primary and secondary A1 components (Schnurr et al. 
2008a). Initial and current stellar mass estimates are shown in 
Table~\ref{ngc3603_params}, and include a (high) 
initial mass of 166 $\pm$ 20 M$_{\odot}$ for NGC 3603 B. 
Independent age estimates using pre-main sequence isochrones of low mass stars also
favour low 1$\pm$1 Myr ages (Sung \& Bessell 2004), while Crowther
et al. (2006) estimated 1.3$\pm$0.3 Myr for NGC 3603 from a comparison between
massive O stars and theoretical isochrones (Lejeune \& Schaerer 2001).

\subsection{R136}

Fig.~\ref{r136_fig1to4} compares the derived properties of R136a1, a2, a3 
and c with LMC metallicity evolutionary predictions, under the assumption 
that these stars are single. Initial stellar masses in the 
range 165 -- 320 M$_{\odot}$ are implied, at ages of 1.7 $\pm$ 0.2 Myr, 
plus high initial rotational rates, in order that the observed surface 
hydrogen contents of 30 -- 40\% by mass are reproduced. 
Initial and current
stellar mass estimates are included in Table~\ref{r136_summary}. 
Differences in age 
estimates reflect variations in initial rotation rates. Nevertheless, 
equatorial rotation rates of $v_{e} \sim$ 200 (300) km s$^{-1}$ are predicted 
after $\sim$1.75 Myr (2.75 Myr) for a 300 (150) M$_{\odot}$ star. 

\begin{figure}
\begin{center}
\includegraphics[width=0.6\columnwidth,angle=270,clip]{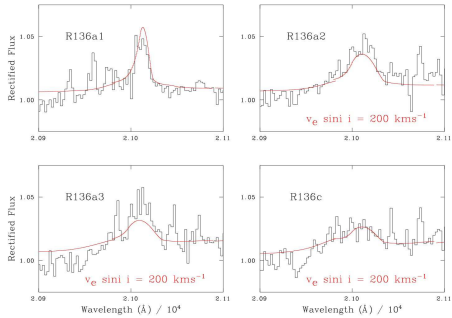}
\caption{Spectral comparison between VLT/SINFONI spectroscopy of 
N\,{\sc 
v} 2.10$\mu$m in R136 WN5 stars (Schnurr et al. 2009) and synthetic 
spectra (red), allowing for instrumental broadening plus rotational
broadening of 200 km\,s$^{-1}$ for R136a2, a3 and c.}
\label{r136_nv}
\end{center}
\end{figure}

In the absence of photospheric absorption features, the N\,{\sc v} 
2.100$\mu$m feature provides the best diagnostic of rotational broadening 
for the R136 stars. This recombination line is intrinsically narrow 
at the SINFONI spectral resolution (FWHM $\sim$ 15\AA)  because it is 
formed extremely deep in the stellar wind. Indeed,
FWHM $\sim$ 15\AA\ is observed for R136a1, as expected either for a 
non-rotating star, or one viewed pole-on (see Fig.~\ref{r136_nv}).
In  contrast, a2 and a3 reveal FWHM $\sim$ 40\AA, corresponding 
to $v_{e} \sin i \approx$ 200 km\,s$^{-1}$. We are unable to 
quantify FWHM  for R136c since the N\,{\sc v} feature is very weak, 
although it too is  consistent with a large rotation rate. Therefore, a2, 
a3 and probably c show spectroscopic evidence for rapid rotation  as 
shown in Fig.~\ref{r136_nv}, while  a1 could either be a very slow 
rotator, or a rapid rotator viewed close to pole-on. 

For comparison, Wolff et al. (2008) found that R136 lacks
slow rotators among lower mass 6--30  $M_{\odot}$ stars. 
Wolff et al. derived $v_{e} \sin i$ = 189 $\pm$ 23 km\,s$^{-1}$ for  
eleven 15--30 $M_{\odot}$  stars within R136 versus $v_{e} \sin i$ = 129 
$\pm$ 13 km\,s$^{-1}$ from equivalent mass field stars within the LMC.
A much more extensive study of rotational velocities for O stars in 30 
Doradus will be provided by the VLT-FLAMES Tarantula Survey (Evans et al. 
2010a).

Finally, although we defer the possibility that
the WN5 stars in R136 are (equal mass) binaries until Sect. 5, 
it is necessary to remark upon the possibility of 
chance superpositions within the observed cluster. 
Line-of-sight effects should not play a significant role in
our interpretation of bright systems such as R136a1 (Ma\'{i}z Apell\'{a}niz 2008).
Chance superposition of other stars has been calculated to contribute at most 
$\sim$10 to 20\% of the (visible)  light (J. Ma\'{i}z Apell\'{a}niz, priv. comm.). 
For example, a 0.2 mag decrease in the  absolute K-band magnitude of R136a1 arising from 
the contribution of lower mass stars along this sightline would lead to a 
10\% reduction in its initial (current) stellar mass, i.e. to 285 M$_{\odot}$ (235 $M_{\odot}$).


\subsection{Synthetic spectra}

\begin{figure*}
\begin{center}
\includegraphics[width=1.0\columnwidth,angle=270,clip]{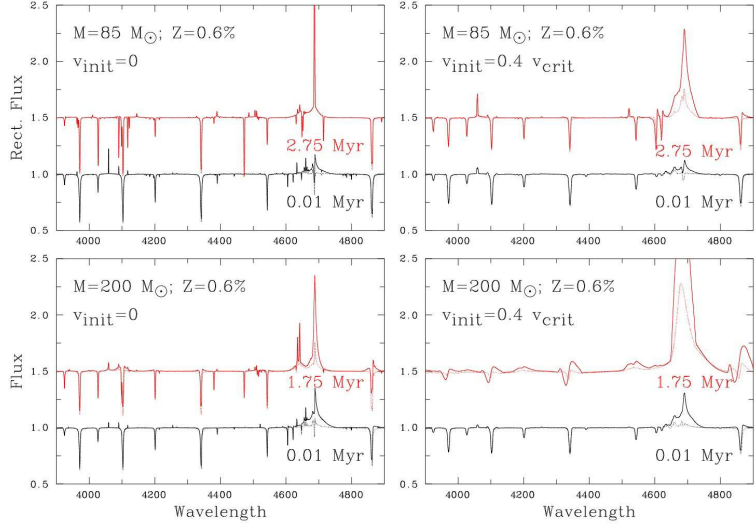}
\caption{Synthetic visual spectra calculated for zero-age main-sequence 
stars (black) of initial mass 85 and 200 M$_{\odot}$, 
initially rotating at $v_{\rm init}/v_{\rm crit}$  = 0 or 0.4,
assuming that winds are clumped ($f_{\infty}$ = 0.1, solid lines) and follow Vink et al. (2001).
Synthetic spectra are also shown at the time at which X$_{\rm H}$ = 30\% by mass 
is predicted at the surface of the initially rotating model, i.e. 1.75 - 2.75 Myr (red).
Synthetic spectra are broadened by 50 km\,s$^{-1}$ ($v_{\rm init}/v_{\rm crit}$  
= 0), $v_{e} \sin i $ = 250 km\,s$^{-1}$ ($v_{\rm init}/v_{\rm crit}$ = 0.4, 
0.01 Myr), $v_{e} \sin i $ = 150 
km\,s$^{-1}$ ($v_{\rm init}/v_{\rm crit}$ = 0.4, 1.75--2.75 Myr). 
He\,{\sc ii} $\lambda$4686 emission is predicted for all phases for these clumpy winds -- 
most prominently during the later phases of initially rapidly rotating very massive stars.
If we were to adopt non-clumped winds ($f_{\infty}$ = 1, dotted lines), He\,{\sc ii} $\lambda$4686 
absorption would be predicted for the ZAMS 85 M$_{\odot}$ models.}
\label{synthetic}
\end{center}
\end{figure*}

We have calculated synthetic spectra for each of the LMC metallicity 
models, both at the Zero Age Main Sequence (ZAMS) and ages corresponding 
to the surface hydrogen compositions reaching X$_{\rm H}$ = 30\% for the
rotating models. These are presented in Fig.~\ref{synthetic} in which
mass-loss rates follow the  theoretical mass-loss recipes 
from Vink et al. (2001), both for the case of radially-dependent
{\it clumped} winds with a volume filling factor of $f_{\infty}$=0.1 at
$v_{\infty}$ (solid)
and {\it smooth} ($f_{\infty}$=1, dotted) winds. For clumped winds, these result in 
ZAMS synthetic O supergiant spectra 
which are equivalent to ~O3\,If (for 85 M$_{\odot}$) and O2-3\,If/WN5-6 
(for  200 M$_{\odot}$) subtypes. Weaker He\,{\sc ii} $\lambda$4686 emission
would naturally be predicted if we were to instead adopt {\it smooth} winds, including 
O3\,V for the 85 M$_{\odot}$ case and O2\,III--If for the 200 M$_{\odot}$ case at 
the zero age main sequence. 

Although specific details depend upon the degree of wind clumping in O 
star winds and validity of the smooth Vink et al. predictions, stars whose 
masses exceed a given threshold are expected to display a supergiant 
signature from the outset. Overall, these results suggest a paradigm shift 
in our understanding of early O dwarfs, which have hitherto been 
considered to represent ZAMS stars for the {\it highest} mass stars 
(Walborn et al. 2002).

For the NGC 3603 and R136 WN stars studied 
here, wind clumping is required to reproduce the 
electron scattering wings  of He\,{\sc ii} $\lambda$4686 (Hillier 
1991). Overall, we find $\log \dot{M} - \log \dot{M}_{\rm Vink}$ = +0.13 $\pm$ 
0.09, in close agreement with Martins et al. (2008) who 
found that spectroscopic mass-loss rates for the Arches WN7--9 stars
exceeded predictions by +0.2 dex. 

Very powerful winds naturally result from the dependence of the mass-loss 
rate upon the ratio of radiation pressure to gravity, $\Gamma_{e} \propto 
L/M$. Since L $\propto M^{1.7}$ for ZAMS stars in the range 30 -- 
300 M$_{\odot}$ based upon evolutionary calculations discussed above, 
$\Gamma_{e} \propto M^{0.7}$. A 30 
M$_{\odot}$ main-sequence star possesses an Eddington parameter of 
$\Gamma_{e} \sim$  0.12 (Conti et al. 2008), 
therefore a 300 M$_{\odot}$ star will possess $\Gamma_{e} \sim 0.5$. 
The latter achieve $\Gamma_{e} \sim 0.7$ after 1.5 Myr, typical of 
the cluster age under investigation here. $\Gamma_{e}$ increases from
0.4 to 0.55 for stars of initial mass 150 M$_{\odot}$ after 1.5 
Myr.

Therefore, the very highest mass stars in very young 
systems may never exhibit a normal O-type absorption line spectrum. 
Indeed, O2--3\,V stars may rather be limited to somewhat lower ZAMS stars, with
a higher mass threshold at lower metallicity. Recall the relatively low dynamical mass 
of 57  M$_{\odot}$ for the O3 V primary in R136-38 (Massey et al. 2002), which
currently represents the most massive O2--3 V star dynamically weighed, versus
83 M$_{\odot}$ (WR20a: 
Bonanos et al. 2004; Rauw et al. 2005), $\geq$87 
M$_{\odot}$ (WR21a: Niemela et al. 2008)  and 116 M$_{\odot}$ (NGC 
3603 A1a: Schnurr et al. 2008a) for some of the most massive H-rich WN 
stars. O\,If$^{\ast}$  or even WN-type emission line spectra may be expected for 
the highest mass  stars within the very youngest Giant H\,{\sc ii} regions, as is 
the case for  R136 and NGC 3603 here, plus Car OB1 (Smith 2006) and W43 
(Blum et al. 1999).

\section{Binarity}

NGC 3603 A1 and C are  confirmed binaries (Schnurr et al. 2008a),  and 
R136c is a probable binary (Schnurr et al. 2009) while component B and 
R136a1, a2, a3 are presumed to be single. Of course, we cannot 
unambiguously confirm that  these are genuinely single but
if their mass ratios differ greatly from unity the derived physical
properties reflect those of the primary. Therefore, here we shall
focus upon the possibility that the R136 stars are massive 
binaries, whose ratio is close to unity (Moffat 2008). 

If we adopt similar mean molecular weights, $\mu$, for individual components, 
current 
(evolutionary) masses of 150 + 150 M$_{\odot}$, 200 + 100 M$_{\odot}$ and 
220 + 55 M$_{\odot}$ would be required to match the current properties of 
R136a1 for mass ratios of 1, 0.5 and 0.25 since L $\propto \mu M^{1.5}$. 
Of these possibilities, only near-equal mass binaries would contradict
the high stellar masses inferred here. 
If the separation between putative binary components were small, radial 
velocity variations would be expected. There is no  unambiguous 
evidence for binarity among the stars under investigation, although R136c 
does exhibit marginal radial velocity variability (Schnurr et al. 2009). 

In contrast, the WN6h system A1 in the young Milky Way cluster NGC 3603 
{\it is }
a short-period massive binary (Schnurr et al. 2008a). However, longer 
period systems with periods of months to years could have easily avoided 
spectroscopic detection.  We shall employ anticipated properties of 
colliding wind systems and dynamical interactions to assess whether the 
R136a stars are realistically long-period binaries. 
This approach is especially sensitive  to equal wind momenta (equal mass)
systems since the X-ray emission is maximal 
for the case of winds whose momenta are equal. A1 in NGC 3603 is 
relatively faint in X-rays because the components are in such close 
proximity that their winds collide at significantly below maximum 
velocity.

\begin{table}
\begin{center}
\caption{Predicted X-ray luminosities from R136a1 for scenarios in 
which it is either a colliding wind binary using analytical predictions 
from Stevens et al. (1992) and Pittard \& Stevens (2002), or 
single/multiple for various empirical  $L_{\rm X}/L_{\rm Bol}$
values. Analytical calculations adopt component
separations of $d$ = 3, 30, 300 AU, for mass ratios: (i) $q$ = 1  (150 + 150 $M_{\odot}$) and 
equal wind momenta, $\eta$ = 1  ($\dot{M}_{1} = 2.8 \times 10^{-5}$  $M_{\odot}$\,yr$^{-1}$, 
$v_{\infty, 1}$ = 2600 km\,s$^{-1}$); (ii) $q$ = 0.25 (220 + 55 $M_{\odot}$) and $\eta$ = 0.05
($\dot{M}_{1} = 4.2 \times 10^{-5}$  $M_{\odot}$\,yr$^{-1}$, 
$v_{\infty, 1}$ = 2600 km\,s$^{-1}$, 
$\dot{M}_{2} = 3 \times 10^{-6}$  $M_{\odot}$\,yr$^{-1}$,  
$v_{\infty, 2}$ = 2000 km\,s$^{-1}$). 
The intrinsic X-ray luminosity of
R136a is taken from Guerrero \& Chu (2008).} 
\label{r136a1_xray} 
\begin{tabular}{l@{\hspace{1.5mm}}l@{\hspace{1.5mm}}l@{\hspace{0.5mm}}r@{\hspace{1.5mm}}r@{\hspace{1.5mm}}r
@{\hspace{2mm}}r@{\hspace{2mm}}r@{\hspace{2mm}}r@{\hspace{2mm}}r}
\hline 
Source & Sp  & $q$ & $d$ & Period & $\chi_{1}$ & $\Xi_{1}$ & $\chi_{2}$ & $\Xi_{2}$ & $L_{\rm X, 1+2}$ \\
       & Type&     & AU  & year   &            &          &             &           & 10$^{34}$ \\
       &     &     &     &        &            &          &             &           & erg\,s$^{-1}$ \\
\hline
R136a1 & 2$\times$WN5 &1.0 & 3          & 0.3    &   3.7            &  0.17   & 3.7   & 0.17 & 540\phantom{.3}   \\
R136a1 & 2$\times$WN5&1.0 & 30         & 9.5    &   37\phantom{.7} &  0.17   & 37\phantom{.7} & 0.17 &   54\phantom{.3}   \\
R136a1 & 2$\times$WN5&1.0 & 300 & 300\phantom{.5} & 370\phantom{.7}&  0.17   & 370\phantom{.7} & 0.17 & 5.4   \\
R136a1 & WN5+O&0.25 & 3              & 0.3 &   3.7             & 0.02   & 4.5   & 0.44 & 88\phantom{.3}\\
R136a1 & WN5+O&0.25 & 30             & 9.8 &   37\phantom{.7}  & 0.02   & 45\phantom{.7} & 0.44 & 8.8 \\
R136a1 & WN5+O&0.25 & 300 & 310\phantom{.5}&   370\phantom{.7} & 0.02   & 450\phantom{.7}& 0.44   & 0.9 \\
\hline
R136a1 & WN5 & \multicolumn{7}{l}{Single, $L_{\rm X}/L_{\rm Bol} = 10^{-7}$} & 0.3  \\
R136a1 & WN5+O? & \multicolumn{7}{l}{Binary, $L_{\rm X}/L_{\rm Bol} = 10^{-6}$} & 3.3  \\
R136a1 & 2$\times$WN5? & \multicolumn{7}{l}{Binary, $L_{\rm X}/L_{\rm Bol} = 5 \times 10^{-6}$} & 16.3  \\
\hline
R136a  & &     &     &                &                   &        &                &        & 2.4 \\
\hline
\end{tabular}
\end{center}
\end{table}

\subsection{X-rays}

If both components in a binary  system possessed similar masses, once 
outflow velocities of their dense  stellar winds achieve asymptotic 
values, collisions would produce stronger  X-ray emission 
than would be expected from a single star. Analytical
estimates of X-ray emission from colliding wind systems are approximate
(Stevens et al. 1992).  Nevertheless, this approach does provide 
constraints upon orbital separations to which other techniques are 
currently insensitive and is especially sensitive  to systems whose wind
strengths are equal. According to Pittard \& Stevens (2002) up to 17\% of 
the wind power 
of the primary (and 17\% of the secondary) can be radiated in X-rays for 
equal wind momenta systems, versus 0.4\% of the primary wind power (56\% 
of the secondary) for (unequal mass) systems whose wind momentum ratio is 
$\eta$ = 0.01.

To illustrate the diagnostic potential for X-ray observations, let us 
consider the expected X-ray emission from R136c under the 
assumption that it is single.  The
intrinsic X-ray luminosities of single stars can be approximated by
L$_{\rm X}$/L$_{\rm Bol} \sim 10^{-7}$ (Chlebowski et al. 1989).
From our spectroscopic analysis we would
expect L$_{\rm X} \approx 2 \times 10^{33}$ erg\,s$^{-1}$, yet Chandra
imaging reveals an intrinsic X-ray luminosity which is a factor of 30--50
times higher (Portegies Zwart et al. 2002; Townsley et al. 2006; Guerrero 
\& Chu 2008), arguing for a colliding wind system in this 
instance. Analytical models favour equal mass components separated by 
$\sim$100 Astronomical Units (AU).

In contrast, the expected X-ray emission from the sum of R136a1, a2 
a3 (and a5) -- unresolved at Chandra resolution -- is L$_{\rm X} \approx 7
\times 10^{33}$ erg\,s$^{-1}$ under the assumption that they are single.
In this case, the intrinsic X-ray emission from R136a is observed to be 
only a factor of $\sim$3 
higher (Portegies Zwart et al. 2002; Guerrero \& Chu 2008) arguing against
short period colliding wind systems from any R136a components. 
Multiple wind interactions (outside of the binaries but within 
the  cluster) will also produce shocks and X-ray emission (see e.g.
Reyes-Iturbide et al. 2009).

Empirically, colliding winds within O-type binaries typically exhibit 
L$_{\rm X}$/L$_{\rm Bol} \sim 10^{-6}$ (Rauw et al. 2002), although 
binaries comprising stars with more powerful winds (i.e. Wolf-Rayet stars) 
often possess stronger X-ray emission. Indeed, NGC~3603 C has an X-ray 
luminosity of $L_{\rm X} \gtsim 4 \times 10^{34}$ erg\,s$^{-1}$ (Moffat et 
al. 2002), corresponding to L$_{\rm X}$/L$_{\rm Bol} \gtsim 5 \times 
10^{-6}$ based upon our spectroscopically derived luminosity (similar 
results are obtained for R136c). If we were to assume this ratio for the 
brightest components of R136a (a1, a2, a3 and a5) under the assumption 
that each were close binaries -- i.e. disregarding predictions from 
colliding wind theory -- we would expect an X-ray luminosity that is a 
factor of 15 times higher than the observed value.
Therefore, the WN stars in R136a appear to be single, possess 
relatively low-mass companions or have wide separations.

If we make  the reasonable assumption that equal mass stars would possess 
similar  wind  properties, we can consider the effect of wind collisions 
upon the  production of X-rays. Let us consider a 150 + 150 
M$_{\odot}$ binary  system  with a period of 100 days in a circular orbit 
with separation 3  AU, whose individual components each possess mass-loss 
rates of $\dot{M}$  = 2.8 $\times$ 10$^{-5}$ M$_{\odot}$ yr$^{-1}$ and 
wind 
velocities of  $v_{\infty}$ = 2600 km s$^{-1}$. A pair of stars with 
such properties 
could match the appearance of R136a1, with individual properties fairly 
representative of R136a3. 


Since the ratio of their wind momenta, $\eta$ is unity, 
the fraction, $\Xi$, of 
the wind kinetic power  processed in the shock from each star
is maximal  ($\Xi \sim$1/6, Pittard \& Stevens 2002). One must calculate the 
conversion efficiency, $\chi$, of 
kinetic wind power into radiation for each star. From Stevens et al. (1992)
\[ \chi \approx \frac{v_{8}^{4} d_{12}}{\dot{M}_{-7}}, \]
where $v_{8}$ is the wind velocity in units of 10$^{8}$ cm\,s$^{-1}$, 
$d_{12}$ is 
the distance to the interaction region  in units of 10$^{12}$ cm (i.e. 
half the separation for equal wind momenta) and $\dot{M}_{-7}$ is 
the mass-loss rate in units of 10$^{-7} M_{\odot}$\,yr$^{-1}$.  We set a 
lower limit of  $\chi$ = 1 for cases in which the system radiates all of 
the  collision energy. This allows an estimate of the intrinsic X-ray 
luminosity (J. Pittard, priv. comm.),
\[
L_{\rm X} \approx \frac{1}{2} \frac{\dot{M} v_{\infty}^2 \Xi}{\chi}.
\]
For a 150 + 150 M$_{\odot}$ system with separation of 3 
AU, we derive $\chi \sim$ 3.7 and  predict L$_{\rm X}$ = 2.7  $\times$ 
10$^{36}$ erg  s$^{-1}$ for {\it each} component. 
The total exceeds the measured X-ray  luminosity of 
R136a by a factor of $\sim$200 (Townsley et al. 2006; Guerrero \& Chu 
2008), as shown in Table~\ref{r136a1_xray}. Recall that if R136a1 were to
be a massive binary with a mass ratio of 0.25, the secondary would possess 
a mass of $\sim$55 M$_{\odot}$. This would contribute less than 10\% of the observed 
spectrum, with a secondary to primary wind momentum ratio of $\eta\sim$0.05.
Predictions for this scenario are also included in Table~\ref{r136a1_xray}, 
revealing X-ray luminosities a factor of $\sim$6 lower than the equal wind
(equal mass) case.

Stevens et al. (1992) also provide an expression for the 
characteristic intrinsic column density in colliding wind binaries, 
$\bar{N}_{\rm H}$ 
(their Eqn. 11), from which $\bar{N}_{\rm H}$ $\sim$ 2.4 $\times 10^{22}$
cm$^{-2}$ would be inferred for R136a1 is it were an equal mass binary system, 
separated by 3 AU. This is an order of magnitude higher than the
estimate for R136a (Townsley et al. 2006; Guerrero \& Chu 2008), 
arising from a combination of foreground and internal components.

Of course, a significant fraction of the shock energy could produce 
relativistic particles rather than X-ray emission (e.g. Pittard \& 
Dougherty 2006). 
Formally, an equal mass binary system whose components are separated by 
600 AU is
predicted to produce an X-ray luminosity of L$_{\rm X}$ = 2.7 $\times$ 10$^{34}$ erg 
s$^{-1}$, which is comparable to the total intrinsic X-ray luminosity of 
R136a (Guerrero \& Chu 2008). If 90\% of the shock energy were to 
contribute to relativistic particles, the same X-ray luminosity could
arise from a system whose components were separated by 60 AU.
Still, the relatively low X-ray luminosity of R136a suggests that 
if any of the R136a WN5 stars are composed of equal mass binaries, their 
separations would need to be {\it in excess of} 100--200 AU. 
How wide could  very  massive binaries be within such a dense cluster? 
To address this question we now consider dynamical interactions.

\subsection{Dynamical interactions}

The binding energy of a `hard' 100 
M$_{\odot}$ + 100 M$_{\odot}$ binary with a separation of 100 AU is 
comparable to the binding energy of a massive cluster (~10$^{41}$ J). 
Such a system will have frequent encounters with other massive 
stars or binaries. The collision rate for a system of separation, $a$, in 
a 
cluster of number density, $n$, and velocity dispersion, $\sigma$, is 
\[
T_{\rm coll} \sim 30 n \sigma a^{2} (1 - \theta) 
\]
where $\theta$ is  the Safranov number, indicating the importance of 
gravitational focusing  (Binney \& Tremaine 1987). Obviously, an encounter 
with a low mass star  will not affect a very massive binary, so we assume 
only that encounters  with stars in excess of ~50 M$_{\odot}$ will 
significantly affect a very  massive binary. 

For a 100 + 100 M$_{\odot}$ binary with 100 AU separation 
in a cluster with a velocity dispersion of $\sigma$ = 5 km s$^{-1}$, 
initially containing a number density of 500 pc$^{-3}$ ($n$ = 1.5 x 
10$^{-47}$ m$^{-3}$) then $\theta$ = 35. Each very massive binary 
should have an encounter every 1.8 Myr. For binaries wider than 100 AU the 
encounter rate will be even higher, since the encounter rate scales with 
the square of the separation, i.e. encounters every $\sim$0.2 Myr for 
separations of 300 AU or 0.05 Myr for separations of 600 AU.

Such close encounters will create an unstable 
multiple system that will rapidly decay by ejection of the lowest mass 
star (Anosova 1986). This would not necessarily destroy the binary. 
Instead, it will harden the system in order to gain the energy required to 
eject the other star. Thus very massive systems separated by more than 
$\sim$100 AU will reduce their separations through dynamical interactions 
with  other high mass stars within the dense core of R136a.

\begin{figure*}
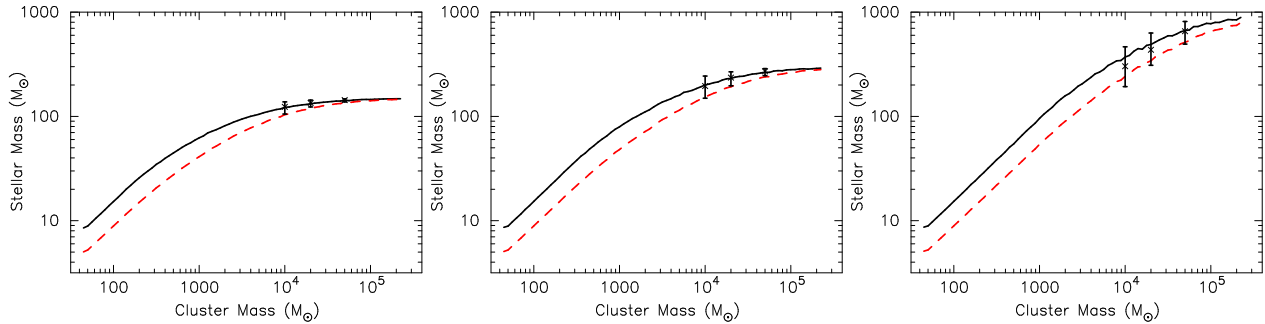

\begin{center}
\includegraphics[width=0.5\columnwidth,angle=270,clip]{cmmsm_av_150topimf_NEW.ps}
\includegraphics[width=0.5\columnwidth,angle=270,clip]{cmmsm_av_300topimf_NEW.ps}
\includegraphics[width=0.5\columnwidth,angle=270,clip]{cmmsm_av_1000topimf_NEW.ps}
\caption{Average mass of the highest mass star (solid black line) and 
25\% and 75\% quartiles for 10$^{4}$, 2  $\times$ 10$^{4}$ and 5 $\times$ 
10$^{4}$ M$_{\odot}$ (vertical error bars), 
plus the average of the three  highest  mass stars (dotted red line) 
versus cluster mass. Clusters were randomly populated by  sampling a 
three-part IMF  (Kroupa 2002) using a Monte Carlo simulation, under the 
assumption that the upper mass limit is (left) 150 M$_{\odot}$; (middle) 
300 M$_{\odot}$; (right) 1000 M$_{\odot}$.  In order to minimise sampling  
problems we take an `average'  cluster as the mean of 1000 realisations of 
a cluster of a particular mass.}
\label{most_massive_star}
\end{center}
\end{figure*}

Should one of the high mass binaries in R136a elude hardening in the way 
outlined above, would it still escape detection if it remained at a large
separation? One binary with components of 150 + 150 
M$_{\odot}$ separated by $\sim$300 AU for which $\sim$30\% of the
shock energy contributes to the X-ray luminosity, together 
with intrinsic X-ray 
luminosities  from  single R136a  members, obeying L$_{\rm 
X}$/L$_{\rm Bol}$ $\sim 10^{-7}$  (Chlebowski et  al. 1989)  would indeed 
mimic the observed R136a X-ray luminosity (Guerrero \& Chu 2008).

In summary, if we adopt similar ratios of X-ray to bolometric luminosities 
as for R136c and NGC 3603C, we would expect an X-ray luminosity from R136a 
that is a factor of 15 times higher than the observed value if a1, a2 and 
a3 were each colliding wind systems. Alternatively, we have followed the 
colliding wind theory of Stevens et al. (1992) and Pittard \& Stevens 
(2002), and conservatively assume 30\% of the shock energy 
contributes to the X-ray luminosity. If {\it all} the very massive stars 
within the Chandra field-of-view (R136a1, a2, a3 and a5) were members of 
equal mass binaries with separations of 300 AU, we would expect a X-ray 
luminosity that is over a factor of 2 higher than observed. In view of 
dynamical effects, at most one of the WN components of R136a might be a 
long period, large separation ($\geq$300 AU) equal mass binary.  We 
cannot rule out short-period, highly unequal-mass binary systems of 
course, but such cases would have little bearing upon our derived stellar 
masses.


\begin{table}
\begin{center}
\caption{Comparison between the properties of the 
NGC 3603, Arches and R136 star clusters and
their most massive stars, $M_{\rm Max}$.}
\label{clusters}
\begin{tabular}{l
@{\hspace{2mm}}r
@{\hspace{2mm}}c
@{\hspace{2mm}}c
@{\hspace{2mm}}c
@{\hspace{2mm}}c
@{\hspace{2mm}}c
}
\hline 
Name & $M_{\rm cl}$ & $\tau$ & Ref & M$_{\rm Max}^{\rm init}$ & $M_{\rm Max}^{\rm current}$ & Ref\\
     & $M_{\odot}$  & Myr    &     & $M_{\odot}$              & $M_{\odot}$                 & \\
\hline
NGC 3603 & 10$^{4}$ & $\sim$1.5    & a, b & 166$\pm$20 & 132$\pm$13 & b \\
Arches   & 2$\times 10^{4}$ & 2.5$\pm$0.5 & c, d & 120--150 & $\geq$95--120 & b, d \\
         &                  & $\sim$1.8            &  b    & 
$\geq$185$^{+75}_{-45}$ & 
$\geq$130$^{+45}_{-25}$  & b\\
R136     & $\leq$5.5$\times 10^{4}$ & $\sim$1.7 & e, b & 
320$^{+100}_{-40}$ & 265$^{+80}_{-35}$ & b \\
\hline
\end{tabular}
\end{center}
\begin{small}
(a) Harayama et al. (2008); (b) This work; (c) Figer (2008); (d) Martins et al. (2008); (e) Hunter et al. (1995)
\end{small}
\end{table}

\section{Cluster Simulations}

We now compare the highest mass stars in R136 and NGC 3603 with cluster 
predictions, randomly sampled from the stellar IMF, for a range of stellar 
mass limits. For R136, the stellar mass within a radius $\sim$4.7 pc from 
R136a1 has been inferred for $\geq$2.8 M$_{\odot}$ at 2.0 $\times$ 
10$^{4}$ M$_{\odot}$ (Hunter et al.  1995). If the standard IMF (Kroupa 
2002) is adopted for $<$ 2.8 M$_{\odot}$ a total cluster mass of $\leq$5.5 
$\times$ 10$^{4}$ M$_{\odot}$ is inferred. Andersen et al. (2009) 
estimated a factor of two higher mass by extrapolating their measured mass 
down to 2.1 M$_{\odot}$ within 7 pc with a Salpeter slope to 0.5 
M$_{\odot}$. These are upper limits to the mass of R136 itself since the 
mass function in the very centre of R136 cannot be measured (Ma\'{i}z 
Apell\'{a}niz 2008), such that they include the associated halo of older 
($\geq$ 3 Myr)  early-type stars viewed in projection -- indeed Andersen 
et al. (2009) obtained an age of 3 Myr for this larger region. Examples of 
such stars include R136b (WN9h), R134 (WN6(h)) and Melnick 33Sb (WC5).

For NGC 3603, we adopt a cluster mass of 1.0 $\times 10^{4}$ M$_{\odot}$, 
the lower limit obtained by Harayama et al. (2008) from high resolution 
VLT NAOS/CONICA imaging. These are presented in Table~\ref{clusters}, 
together with the Arches cluster, for which Figer (2008) estimated a 
cluster mass of 2 $\times 10^{4}$ M$_{\odot}$ based on the mass function 
of Kim et al. (2006).

\subsection{Cluster populations}

We simulate a population of clusters and stars by randomly sampling first 
from a power-law cluster mass function (CMF), and then populating each 
cluster with stars drawn randomly from a stellar IMF (Parker \& Goodwin 
2007). Cluster masses are selected from a CMF of the form $N(M) \propto 
M^{-\beta}$, with standard slope $\beta$ = 2 (Lada \& Lada 2003) between 
cluster mass limits 50 M$_{\odot}$ and 2 $\times$ 10$^{5}$ M$_{\odot}$. 
These limits enable the full range of cluster masses to be sampled, 
including those with masses similar to that of R136. The total mass of 
clusters is set to 10$^{9}$ M$_{\odot}$ to fully sample the range of 
cluster masses. Each cluster is populated with stars drawn from a 
three-part IMF (Kroupa 2002) of the form

\begin{equation*}
N(M)   \propto  \left\{ \begin{array}{ll} M^{+0.3} \hspace{0.4cm} m_0  < M/{\rm M_{\odot}} < m_1   \,, \\ 
M^{-1.3} \hspace{0.4cm} m_1 < M/{\rm M_{\odot}} < m_2   \,, \\ 
M^{-2.3} \hspace{0.4cm} m_2 < M/{\rm M_{\odot}} < m_3   \,,
\end{array} \right.
\end{equation*}

where $m_0 = 0.02$\,M$_\odot$, $m_1 = 0.1$\,M$_\odot$ and $m_2 = 
0.5$\,M$_\odot$. We use three different values for $m_{3}$ in the 
simulations; 150 M$_{\odot}$, 300 M$_{\odot}$ and 1000 M$_{\odot}$. 
Stellar mass is added to the cluster until the total mass is within 2\% of 
the cluster mass. If the final star added to the cluster exceeds this 
tolerance, then the cluster is entirely repopulated (Goodwin \& Pagel
2005).

Our random sampling of the IMF allows low-mass clusters to be composed of 
one massive star and little other stellar material. This contravenes the 
proposed fundamental cluster mass-maximum stellar mass (CMMSM) relation 
(Weidner \& Kroupa 2006, Weidner et al. 2010). However, the average 
maximum stellar mass for a  given cluster mass closely follows the CMMSM 
relation (Parker \& Goodwin 2007, Maschberger \& 
Clarke 2008). The results for our three Monte Carlo runs are shown in 
Fig~\ref{most_massive_star}. They reveal that the average relation between 
the  maximum stellar mass and cluster mass (in the 10$^{2}$ and 10$^{4}$ 
M$_{\odot}$ interval) is recovered (Weidner et al. 2010), without the 
constraint that the cluster mass governs the maximum possible stellar mass 
(Parker \& Goodwin 2007). We also indicate 25\% and 75\% quartiles for 
the instances of a 10$^{4}$ M$_{\odot}$ (NGC 3603), 2  $\times$ 10$^{4}$ 
M$_{\odot}$ (Arches) and 5 $\times$ 10$^{4}$ M$_{\odot}$ (R136) cluster.

\subsection{R136}

If we assume that R136a1, a2 and c are either single or the primary 
dominates the optical/IR light, we obtain an average of 260 M$_{\odot}$ 
for their initial masses. In reality this value will be an upper limit due 
to binarity and/or line-of-sight effects. Nevertheless, from 
Fig.~\ref{most_massive_star}, we would expect the average of the three 
most massive stars in a cluster of mass 5 $\times 10^{4}$ M$_{\odot}$ to 
be 150 M$_{\odot}$, 230, 500 M$_{\odot}$ for an adopted upper mass limit 
of $m_{3}$ = 150, 300 and 1000 M$_{\odot}$, respectively. Therefore, an 
upper limit close to 300 M$_{\odot}$ is reasonably consistent with the 
stellar masses derived (see also Oey \& Clarke 2005).

In the lower panel of Figure~\ref{mass_function} we therefore present 
typical mass  functions for a cluster of mass 5 $\times 10^{4}$ 
M$_{\odot}$ for an adopted upper limit of m$_{3}$ = 300 M$_{\odot}$. From
a total of 1.05--1.10 $\times$ 10$^{5}$ stars we would expect $\sim$14 
initially more massive than 100 M$_{\odot}$ 
to have formed within 5 parsec of R136a1, since this relates to the radius 
used by Hunter et al. (1995) to derive the cluster mass. 
Indeed, of the 12  brightest near-infrared sources 
within 5 parsec of R136a1, we infer initial evolutionary masses in excess of 100 
M$_{\odot}$ for 10 cases (all  entries in Table~\ref{r136_photom} except 
for R136b and R139). Of course, this comparison should be tempered by (a)
contributions of putative secondaries to the light of the primary 
in case of binarity; (b) dynamical ejection during the formation process
(e.g. Brandl et al. 2007). Indeed, Evans et al. (2010b) propose that
an O2\,III-If$^{\ast}$ star from the Tarantula survey is a potential high 
velocity runaway from R136, in spite of its high stellar mass ($\sim 90 M_{\odot}$).

Overall, R136 favours a factor of $\approx$2 higher stellar mass limit 
than is currently  accepted. We now turn to NGC 3603 to assess whether it 
supports or contradicts this result.

\subsection{NGC 3603}

On average, the highest initial stellar mass expected in a 10$^{4}$ 
M$_{\odot}$ cluster would be $\sim$ 120 M$_{\odot}$ with an upper mass 
limit of m$_{3}$ = 150 M$_{\odot}$, $\sim$~200 M$_{\odot}$ for m$_{3}$ = 
300 M$_{\odot}$ and $\sim$ 400 M$_{\odot}$ for m$_{3}$ = 10$^{3}$ 
M$_{\odot}$. Since we infer an initial mass of 166$_{-20}^{+20}$ 
M$_{\odot}$ for component B, a stellar limit intermediate between 150 and 
300 M$_{\odot}$ might be expected, if it is either single or dominated by 
a single component\footnote{We do not attempt to claim that NGC 3603 B is 
single, even though it is photometrically and spectroscopically stable and 
is not X-ray bright (Moffat et al. 2004; Schnurr et al. 2008a).}. However, 
no stars initially more massive than 150 M$_{\odot}$ would be expected in 
25\% of 10$^{4}$ M$_{\odot}$ clusters for the case of m$_{3}$ = 300 
M$_{\odot}$ (recall middle panel of Fig.~\ref{most_massive_star}).

\begin{figure}
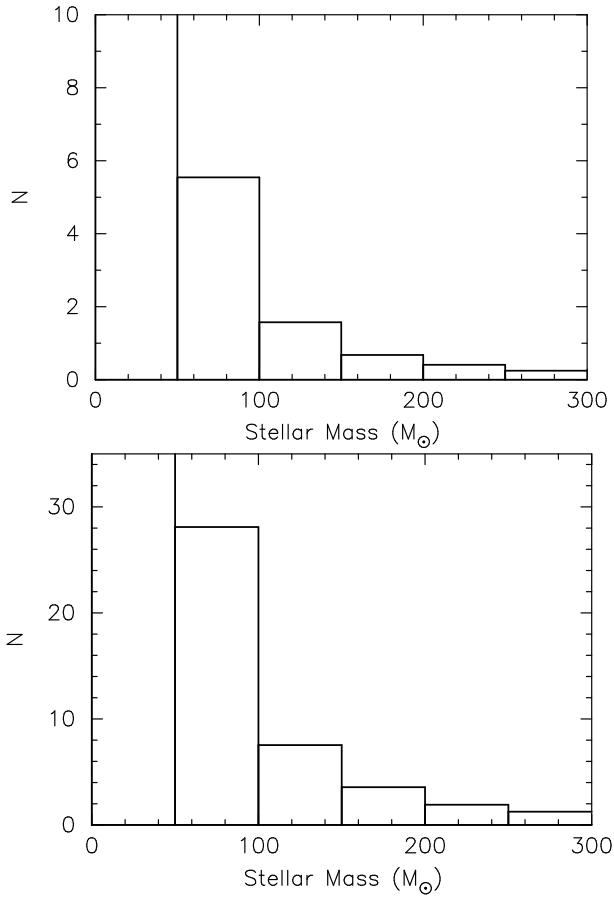

\begin{center}
\includegraphics[width=0.7\columnwidth,angle=270,clip]{mass_dist_104_150_b.ps}
\includegraphics[width=0.7\columnwidth,angle=270,clip]{mass_dist5104_150_b.ps}
\caption{Theoretical  mass functions for clusters with stellar 
masses of (upper) 
10$^{4}$ M$_{\odot}$ and (lower) 5 $\times 10^{4}$ M$_{\odot}$, for the 
scenario with an upper mass limit of $m_{3}$ = 300 M$_{\odot}$. 3 and 14 
stars with initial masses in excess of 100 M$_{\odot}$ are anticipated, 
respectively. Our Monte Carlo approach (Parker \& Goodwin 2007) breaks 
down for very massive stars in the upper panel due to small number statistics.}
\label{mass_function}
\end{center}
\end{figure}

In the upper panel of Figure~\ref{mass_function} we present typical mass 
functions for a cluster of mass 1 $\times 10^{4}$ M$_{\odot}$ for an 
adopted upper mass limit of m$_{3}$ = 300 M$_{\odot}$. We would expect 3 
stars initially more massive than 100 M$_{\odot}$, versus 4 observed in 
NGC 3603 (Table~\ref{ngc3603_photom}). If we were to extend the upper 
limit of the IMF to m$_{3}$ = 1000 M$_{\odot}$, then the average of the 
three most massive stars of a 10$^{4}$ M$_{\odot}$ cluster would be 230 
M$_{\odot}$ which is not supported by NGC 3603 (150 M$_{\odot}$).

Therefore, the very massive stars inferred here for both R136 and NGC 3603 
are fully consistent with a mass limit close to m$_{3}$ = 300 M$_{\odot}$, 
both in terms of the number of stars initially exceeding 100 M$_{\odot}$ 
and the most massive star itself. Before we conclude this section, let us 
now consider the Arches cluster.

\begin{table*}
\begin{center}
\caption{Comparison between stellar properties (and inferred masses) 
of the most luminous
Arches stars from Martins et al. (2008) obtained using different photometry, 
foreground interstellar extinctions and  distances to the Galactic 
Centre.}
\label{arches}
\begin{tabular}{l
@{\hspace{2mm}}l
@{\hspace{2mm}}l
@{\hspace{2mm}}l
@{\hspace{2mm}}c
@{\hspace{2mm}}c
@{\hspace{2mm}}c
@{\hspace{2mm}}c
@{\hspace{2mm}}c
@{\hspace{2mm}}c
@{\hspace{2mm}}c
@{\hspace{2mm}}c
@{\hspace{2mm}}c
@{\hspace{2mm}}c
@{\hspace{2mm}}c
@{\hspace{2mm}}c
}
\hline 
Name & Sp Type & m$_{\rm K_{s}}$& m$_{\rm H}$ - m$_{\rm K_{s}}$ & Ref & 
A$_{\rm K_{s}}$ & Ref & (m-M)$_{0}$ & Ref & M$_{\rm K_{s}}$ & BC$_{\rm 
K_{s}}^{\ast}$ & Ref & $\log L$ & M$_{\rm init}$ & 
M$_{\rm current}$ & Ref\\
  &    & mag & mag & & mag & & mag &  & mag & mag &  & $L_{\odot}$ & M$_{\odot}$ & M$_{\odot}$ & \\
\hline
F6 & WN8--9h & 10.37$^{\ddag}$ & 1.68$^{\ddag}$ & a & 2.8 $\pm$ 0.1 & b & 
14.4 $\pm$ 0.1 & c & --6.83 $\pm$ 0.14 & --4.0 $\pm$ 0.35 & d & 6.25 $\pm$ 
0.15 & $\geq$115$^{+30}_{-25}$ & $\geq$\phantom{0}90$^{+17}_{-13}$ & h\\
   &       & 10.07 & 1.96 & e & 3.1 $\pm$ 0.2 & f & 14.5 $\pm$ 0.1 & g & 
--7.53 $\pm$ 0.22 & --4.0 $\pm$ 0.35 & d & 6.53 $\pm$ 0.16 & 
$\geq$185$^{+75}_{-45}$  & $\geq$130$^{+45}_{-25}$ &h\\
F9 & WN8--9h & 10.77$^{\ddag}$ & 1.67$^{\ddag}$ & a & 2.8 $\pm$ 0.1 & b & 
14.4 $\pm$ 0.1& c  & --6.43 $\pm$ 0.14 & --4.4 $\pm$ 0.35 & d & 6.25 $\pm$ 
0.15 &$\geq$115$^{+30}_{-25}$  & $\geq$\phantom{0}92$^{+19}_{-14}$ & h\\
  &        & 10.62 & 1.78 & e & 3.1 $\pm$ 0.2& f & 14.5 $\pm$ 0.1 & g & 
--6.98 $\pm$ 0.22  & --4.4 $\pm$ 0.35 & d & 6.47 $\pm$ 0.16 & 
$\geq$165$^{+60}_{-40}$ & $\geq$120$^{+35}_{-25}$ &h\\
\hline
\end{tabular}
\end{center}
\begin{small}
(a) Figer et al. (2002); (b) Stolte et al. (2002); (c) Eisenhauer et al. (2005); (d) Martins et al. (2008); (e) Espinoza et al. (2009);
(f) Kim et al. (2006); (g) Reid (1993); (h) This work\\
$\ddag$: NICMOS F205W and F160W filters; $\ast$: Includes --0.25 mag 
offset to bolometric correction relative to F. 
Martins (priv. comm.), as reported in Clark et al. (2009).
\end{small}
\end{table*}

\subsection{Arches}

The upper mass limit of $\sim$150 M$_{\odot}$ from Figer (2005) was 
obtained for the Arches cluster from photometry. This was
confirmed from  spectroscopic analysis by Martins et al. (2008) who estimated 
initial masses of 120 -- 150 M$_{\odot}$ range for the most luminous
(late WN) stars.
How can these observations be reconciled with our results for NGC 3603
and R136?  

Let us consider each of the following scenarios:
\begin{enumerate}
\item the Arches cluster is sufficiently old  (2.5 $\pm$ 0.5 Myr,
Martins et al. 2008) that the highest mass stars have already undergone 
core-collapse,
\item  the Arches is unusually deficient in very massive
stars for such a high mass cluster, or
\item Previous studies have underestimated the masses of the highest mass
stars in the Arches cluster.
\end{enumerate}

We have compared the derived 
properties for the two most luminous WN stars in the Arches cluster (F6 and F9)
from Martins et al. (2008) to the solar 
metallicity grids from Sect. 4. These are presented in Table~\ref{arches},
revealing initial masses of $\geq$115$^{+30}_{-25}$ $M_{\odot}$
from comparison with non-rotating models. 
Uncertainties from solely K-band spectroscopy are significantly higher
than our (UV)/optical/K-band analysis of WN stars in NGC 3603 and R136.
Masses represent lower limits since Martins et al. (2008) indicated 
that the metallicity of the Arches cluster is moderately super-solar.

Core hydrogen exhaustion would be predicted to occur after 
$\sim$2.5 Myr for the non-rotating 150 M$_{\odot}$ solar metallicity 
models, with core-collapse SN anticipated prior to an age of 3 Myr. 
However, comparison between the properties of F6 and F9 -- the most 
luminous WN stars in the  Arches cluster -- with our solar metallicity 
evolutionary models suggest ages of $\sim$2 Myr. On this basis, it
would appear that scenario (i) is highly unlikely.

Let us therefore turn to scenario (ii), on the basis that the initial
mass of the most massive star within the Arches cluster was 150 $M_{\odot}$.
On average, the highest mass star in a 2 $\times 10^{4}$ M$_{\odot}$ 
cluster would be expected to possess an initial mass in excess of 200 
$M_{\odot}$ for an upper mass limit of $m_{3}$ = 300 $M_{\odot}$. 
From our simulations, the highest mass star within such a cluster 
spans a fairly wide range (recall  
Fig.~\ref{most_massive_star}). However, an upper limit of 120 (150) $M_{\odot}$ 
would be  anticipated in only 1\% (5\%) of cases if we were to
adopt $m_{3}$ = 300 $M_{\odot}$. Could the Arches cluster be such a statistical
oddity?

Our simulations indicate that a 2 
$\times 10^{4}$ M$_{\odot}$ cluster 
would be expected to host {\it six} stars initially more massive than 100 
$M_{\odot}$ for $m_{3}$ = 300 $M_{\odot}$. Our solar metallicity 
evolutionary models combined with spectroscopic results from Martins et 
al. (2008) suggest that the 5 most luminous WN stars (F1, F4, F6, F7, F9), 
with $\log 
L/L_{\odot} \geq$ 6.15, are consistent with initial masses of $\geq$100 
$M_{\odot}$, providing they are single or one component dominates 
their near-IR appearance.

 Finally, let us turn to scenario (iii), namely that the stellar masses 
of the Arches stars have been underestimated to date. Martins et al. 
(2008) based their analysis upon HST/NICMOS photometry from Figer et al. 
(2002) together with a (low) Galactic Centre distance of 7.6 kpc from 
Eisenhauer et al. (2005) plus a (low) foreground extinction of $A_{\rm 
K_{s}}$ = 2.8 mag from Stolte et al. (2002). Let us also consider the 
resulting stellar parameters and mass estimates on the basis of the 
standard Galactic Centre distance of 8 kpc (Reid 1993), more recent 
(higher)  foreground extinction of $A_{\rm K_{s}}$ = 3.1 $\pm$ 0.19 mag 
from Kim et al. (2006), plus VLT/NACO photometric results 
from Espinoza et 
al. (2009)\footnote{Indeed, if we adopt (H--K$_{s})_{0}$ = --0.11 mag 
(Crowther et al. 2006) for F6 and F9, $A_{\rm K_{s}}$ = 3.2 $\pm$ 0.2 mag
is implied from the Espinoza et al. (2009) photometry.}. 
These yield absolute K-band magnitudes that are 
0.55--0.7 mag brighter than Figer et al. (2002), corresponding to 
0.22--0.28 dex higher bolometric luminosities. With respect to the 
hitherto 150 M$_{\odot}$ stellar mass  limit  identified by Figer (2005), 
these absolute magnitude revisions  would conspire to increasing the limit 
to $\geq$200 M$_{\odot}$.

In Table~\ref{arches} we provide the inferred 
properties of F6 and F9 on the basis of these photometric properties plus 
the stellar temperatures and K-band bolometric corrections derived by 
Martins et al. (2008). Our solar metallicity non-rotating models indicate 
initial masses of $\geq$185$^{+75}_{-45}$ M$_{\odot}$ and 
$geq$165$^{+60}_{-40}$ 
M$_{\odot}$  for F6 and F9, respectively. These are lower 
limits to the actual initial masses, in view of the super-solar 
metallicity of the Arches  cluster (Martins et al. 2008). In total, 
$\sim$5  stars are consistent with initial masses in excess of $\simeq$150 
M$_{\odot}$ (those listed  above), plus a further 5 stars for which 
initial masses exceed $\sim$100 M$_{\odot}$ (F3, F8, F12, F14, F15).

Overall, {\it ten} stars initially 
more massive than $\sim$100 would suggest either that the mass of the 
Arches cluster approaches  3 $\times 10^{4}$ M$_{\odot}$ or several cases 
are near-equal mass binaries (Lang et al. 2005, Wang et al. 2006).
As such, we would no longer require 
that the Arches cluster is a statistical oddity, since the highest mass 
star would not be expected to  exceed 200 M$_{\odot}$ in 25\% of 
2 $\times 10^{4}$ M$_{\odot}$ clusters.

In conclusion, we have attempted to reconcile the properties of the 
highest mass stars in Arches cluster with an upper mass limit of $m_{3} = 
300$ M$_{\odot}$. Based upon the Martins et al. (2008) study, comparison 
with evolutionary models suggests that the initial masses of the most 
massive stars are $\geq$115$^{+30}_{-25}$ M$_{\odot}$ with ages of $\sim$2 
Myr. We find that an upper mass of 120 M$_{\odot}$ would be expected in 
only 1\% of 2$\times 10^{4}$ M$_{\odot}$ clusters for which $m_{3} = 300$ 
M$_{\odot}$. However, use of contemporary near-IR photometry and 
foreground extinctions towards the Arches cluster, together with the 
standard 8 kpc Galactic Centre distance reveal an initial mass of 
$\geq$185$^{+75}_{-45}$ M$_{\odot}$ for the most massive star based on 
solar metallicity evolutionary models, with 4--5 stars consistent with 
initial masses $\geq$150 M$_{\odot}$.  An upper mass of 200 M$_{\odot}$ 
would be expected in 25\% of cases. Robust inferences probably await 
direct dynamical mass determinations of its brightest members, should they 
be multiple.


\section{Global properties of R136}

We will re-evaluate the ionizing and mechanical wind power resulting 
from all hot, luminous stars in R136 and NGC 3603 elsewhere (E. Doran
et al. in preparation). Here, we shall consider the role played by the
very massive WN stars in R136. We have updated the properties of 
early-type stars brighter than M$_{\rm V}$ = --4.5 mag within a 
radius of $\sim$5 parsec from R136a1 (Crowther \& Dessart 1998) to take
account of contemporary T$_{\rm eff}$--spectral type  calibrations 
for Galactic stars  (Conti et al. 2008), 
and theoretical mass-loss rates (Vink et al. 2001) for OB 
stars. Until a census of the early-type stars within R136 is 
complete, we adopt O3 subtypes for those stars lacking spectroscopy (Crowther \& Dessart 1998).
Aside from OB stars and  the WN5h stars discussed here, 
we have included the contribution of  other emission-line stars, namely 
O2--3\,If/WN stars for which we adopt identical temperatures to those of 
the WN5 stars, one other WN5h star (Melnick 34), a WN6(h) star (R134) for 
which we estimate T$_{\ast} \approx$ 42,000 K and a WN9h star (R136b) for 
which we estimate T$_{\ast} \approx$ 35,000 K from its HST/FOS and 
VLT/SINFONI spectroscopy.  Finally, a WC5 star (Melnick 33Sb)   lies at a 
projected distance of 2.9 pc for which we adopt similar wind properties 
and ionizing fluxes to single WC4 stars (Crowther et al. 2002).

\begin{figure}
\begin{center}
\includegraphics[width=0.9\columnwidth,clip]{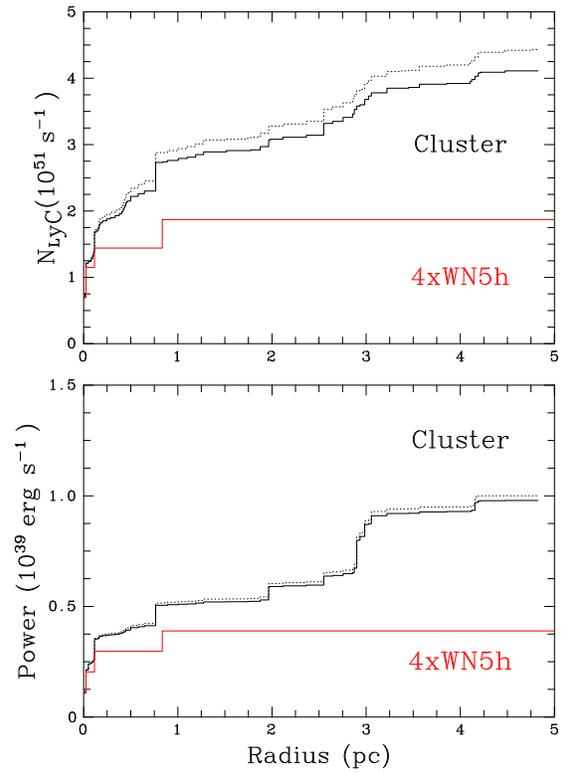}
\caption{(Upper panel) Incremental Lyman continuum ionizing radiation 
from early-type stars within 5 parsec from R136a1 (Crowther \& 
Dessart 1998) either following the 
Galactic O  subtype-temperature calibration (solid black), or 
systematically  increasing the  temperature calibration by 2,000 K for LMC 
stars (dotted line). The  four very  luminous WN5h stars discussed here 
(dashed red line) contribute ~43 - 46\%  of the  total;  (Lower panel) As 
above, except for the mechanical wind power from early-type stars within 
5 parsec of R136a1, based upon theoretical wind prescriptions (Vink et al. 
2001) to which the four WN5h stars contribute ~34\% of the total.}
\label{sum}
\end{center}
\end{figure}

Fig.~\ref{sum} shows the integrated Lyman continuum  ionizing 
fluxes and wind power from all early-type stars in R136, together with the 
explicit contribution of R136a1, a2, a3 and c, amounting to, 
respectively,  ~46\% and 35\% of the cluster total. If we were to adopt a 2,000 K 
systematically higher temperature calibration for O-type stars in the LMC, 
as suggested by some recent results (Mokiem et al. 2007, Massey et 
al. 2009), the contribution of WN5h stars 
to the integrated Lyman continuum flux and wind power is reduced to 43\% 
and 34\%, respectively. 

We have compared our empirical results with 
population synthesis predictions (Leitherer et al. 1999) for a cluster of 
mass 5.5 $\times$ 10$^{4}$ M$_{\odot}$
calculated using a standard IMF (Kroupa 2002) and evolutionary models up 
to a maximum  limit of 120 M$_{\odot}$.
These stars alone approach the mechanical power and ionizing flux 
predicted for a R136-like cluster at an age of $\sim$1.7 Myr (Leitherer et 
al. 1999). Indeed, R136a1 alone provides the Lyman continuum output of seventy 
O7 dwarf stars (Conti et al. 2008), supplying ~7\% of the radio-derived 
$\sim$10$^{52}$  photon s$^{-1}$ ionizing flux from the entire 30 Doradus 
region (Mills et  al. 1978, Israel \& Koornneef 1979). Improved agreement 
with synthesis models
would be expected if evolutionary models allowing for  rotational mixing were used (Vazquez et al. 2007),
together with contemporary mass-loss prescriptions for main-sequence stars.

\section{Discussion and conclusions}

If very massive stars -- exceeding the currently accepted 150 
$M_{\odot}$ limit --  were to exist in the local universe, they would be:
\begin{enumerate}
\item Located in high mass ($\geq 10^{4}$ $M_{\odot}$), very young 
($\leq$ 2 Myr) star clusters;
\item Visually the brightest stars in their host cluster, since L $\propto
M^{1.5}$ for zero age main sequence stars above 85 
M$_{\odot}$\footnote{The exponent flattens further at the highest masses, such that
L $\propto  M^{1.3}$ for zero age main sequence stars between 300--500 
M$_{\odot}$.}, 
and surface
temperatures remain approximately constant for the first 1.5 Myr
\item Possess very powerful stellar winds, as a result of the mass 
dependence of the Eddington parameter, $\Gamma_{e} \propto L/M 
\approx M^{0.5}$ for such stars. 150 -- 300 M$_{\odot}$ zero age main
sequence stars, for which $\Gamma_{e} \approx$ 0.4 -- 0.55, 
would likely possess an O {\it supergiant} morphology, while a Wolf-Rayet
appearance would be likely to develop within the first 1--2 Myr, albeit with 
significant residual surface hydrogen.
\end{enumerate}

In this study we have presented spectroscopic analyses of bright WN 
stars located within R136 in the LMC  and NGC 3603 in the Milky Way 
that perfectly match such  anticipated characteristics. The 
combination of line blanketed  spectroscopic tools  and contemporary 
evolutionary models reveals  excellent agreement with  dynamical mass 
determinations for the components  of A1 in NGC 3603, with  component B 
possessing a higher initial mass of  $\sim$170 M$_{\odot}$,  under the 
assumption that it is single.  Application to the higher temperature 
(see also R\"{u}hling 2008)  brighter members of R136 suggests still 
higher initial masses 
of 165 -- 320 M$_{\odot}$. Owing to their dense stellar winds, they have 
already lost up to 20\% of their initial mass with the first $\sim$1.5 Myr 
of their main sequence evolution (see also de Koter et al. 1998). 
The R136 WN5 stars are moderately hydrogen depleted, to which rotating
models provide the best agreement, whereas the NGC 3603 WN6 stars possess
normal hydrogen contents, as predicted by non-rotating models at early
phases. These differences may arise from different formation mechanisms
(e.g. stellar mergers). Wolff et al. (2008) found a deficit of 
slow rotators within a sample of R136 early-type stars, although
statistically significant results await rotational velocities from the 
VLT-FLAMES Tarantula Survey (Evans et al. 2010a).

We have assessed the potential that each
of the R136 stars represent unresolved equal-mass binary systems. 
SINFONI spectroscopic observations argue against close, short period 
binaries  (Schnurr et al. 2009). R136c possesses an X-ray luminosity
which is a factor of $\sim$100 times greater than that which would be 
expected from a single star and is very likely a massive binary. 
In contrast,  X-ray emission 
from R136a exceeds that expected from single stars by only a factor of 
$\sim$3, to which  multiple wind interactions within the cluster
will also contribute. If the R136a stars possessed similar X-ray properties
to R136c and NGC 3603C, its X-ray emission would be at least a factor
of 15 times higher than the observed luminosity. 
At most, only one of the WN sources within R136a  might be a very long 
period, large separation ($\sim$300 AU), equal-mass 
binary system if as little as 30\% of the shock energy contributes to the
X-ray luminosity. Dynamical effects 
would harden such systems on $\ll$Myr timescales.  We cannot  rule out 
shorter-period, unequal-mass binary  systems of course, but such  cases 
would have little bearing upon our mass limit inferences.

\begin{table}
\begin{center}
\caption{Compilation of stars within R136/30 Dor, NGC~3603 and the Arches 
cluster whose initial masses exceed $\approx$150 M$_{\odot}$ according to 
evolutionary models presented here. Photometry, extinctions and bolometric 
corrections are presented in this study, with the exception of the Arches 
cluster for which we follow Espinoza et al. (2009), Kim et al. (2006) and 
Martins et al. (2008), respectively. Known or suspected binaries are 
marked with  $\ast$.}
\label{vms}
\begin{tabular}{l
@{\hspace{1mm}}l
@{\hspace{1.5mm}}r
@{\hspace{1mm}}c
@{\hspace{1.5mm}}l
@{\hspace{1mm}}l
@{\hspace{1.5mm}}l
@{\hspace{1mm}}l
@{\hspace{1.5mm}}l}
\hline 
Star & Sp Type & m$_{\rm K_{s}}$ & A$_{\rm K_{s}}$ & (m-M)$_{0}$ & M$_{\rm 
K_{s}}$ & BC$_{\rm K_{s}}$ & M$_{\rm Bol}$ & M$_{\rm init}$ \\
     &         & mag             & mag             & mag         & mag             & mag                     & mag           & M$_{\odot}$  \\
\hline
R136a1 & WN5h & 11.1\phantom{5} & 0.2 & 18.45 & --7.6 & --5.0 & --12.6 & 320 \\ 
R136a2 & WN5h & 11.4\phantom{5} & 0.2 & 18.45 & --7.3 & --4.9 & --12.2 & 240 \\
R136c  & WN5h & 11.3\phantom{5} & 0.3 & 18.45 & --7.4 & --4.7 & --12.1 & 220$^{\ast}$ \\
Mk34   & WN5h & 11.7\phantom{5} & 0.3 & 18.45 & --7.1 & --4.8:& --11.9:& 
190$^{\ast}$ \\
Arches F6 & WN8--9h & 10.1\phantom{5} & 3.1 & 14.5 & --7.5 & --4.0 & 
--11.5 & $\geq$185 \\
NGC 3603 B & WN6h & 7.4\phantom{5} & 0.6 & 14.4 & --7.5 & --3.9 & --11.4 & 
166 \\
R136a3 & WN5h & 11.7\phantom{5} & 0.2 & 18.45 & --6.9 & --4.8 & --11.7 & 165 \\
Arches F9 & WN8--9h & 10.6\phantom{5} & 3.1 & 14.5 & --7.0 & --4.4 & 
--11.4 & $\geq$165 \\
Arches F4 & WN7--8h & 10.2\phantom{5} & 3.1 & 14.5 & --7.4 & --3.9 & 
--11.3 & $\geq$150: \\
NGC 3603 & WN6h & 8.0\phantom{5} & 0.6 & 14.4 & --7.0 & --4.2 & --11.2 & 148 \\
      A1a \\
Arches F7 & WN8--9h & 10.3\phantom{5} & 3.1 & 14.5 & --7.3 & --4.0 & 
--11.3 & $\geq$148:\\
Arches F1 & WN8--9h & 10.35 & 3.1 & 14.5 & --7.25 & --4.0 & --11.25 & 
$\geq$145:\\
\hline
\end{tabular}
\end{center}
\begin{small}
\end{small}
\end{table}

In Table~\ref{vms} we present a compilation of stars in R136 (30 Doradus), 
NGC~3603 and the Arches cluster whose initial masses challenge the 
currently accepted upper mass limit of $\sim$150 M$_{\odot}$. Although the 
formation of high mass stars remains an unsolved problem in astrophysics 
(Zinnecker \& Yorke 2007), there are no theoretical arguments in favour of 
such a limit at 150 M$_{\odot}$ (e.g.  Klapp et al. 1987) -- indeed Massey 
\& Hunter (1998) argued against an upper stellar limit based on R136 
itself. Observations of the Arches cluster provide the primary evidence 
for such a sharp mass cutoff (Figer 2005). However, as Table~\ref{vms} 
illustrates, contemporary photometry and foreground extinction towards the 
cluster coupled with the spectroscopy results from Martins et al. (2008) 
suggest 4--5 stars initially exceed $\approx$150 M$_{\odot}$, with an 
estimate of $\geq$185 M$_{\odot}$ for the most luminous star. Recall 
Martins et al. (2008)  used identical spectroscopic tools to those 
employed in the current study, plus near-IR spectroscopic observations 
which also form a central component of our study. On this basis the Arches 
cluster would no longer be a statistical oddity.

Monte Carlo simulations for various upper stellar mass limits 
permit estimates of the revised threshold from which  $\sim$300 
M$_{\odot}$ is obtained. It may be 
significant that both NGC 3603 and R136 are consistent with an identical 
upper limit, in spite of their different  metallicities. Oey \& Clarke 
(2005) obtained an upper limit of $\ll$500 M$_{\odot}$, primarily from a 
maximum          stellar mass of 120 -- 200  M$_{\odot}$ inferred for stars
within R136a itself at that time.

Would there be any impact of an upper mass limit of order 
$\sim$300 M$_{\odot}$ on 
astrophysics? Population synthesis studies are widely applied to 
star-forming regions within galaxies, for which an upper mass limit of 
$\sim$ 120 M$_{\odot}$ is widely adopted (Leitherer et al. 1999). A number of 
properties are obtained from such studies, including star formation rates, 
enrichment of the local interstellar medium (ISM) through mechanical 
energy through winds, chemical enrichment and ionizing fluxes. A 
higher stellar mass limit would increase their global output, especially for 
situations in which rapid rotation leads to stars remaining at high 
stellar temperatures.

Could the presence of very high mass stars, such as those discussed
here, be detected in spatially 
unresolved star clusters? The high luminosities of such stars inherently
leads to the development of very powerful stellar winds at early 
evolutionary phases (1--2 Myr). Therefore the presence of such stars 
ensures that the integrated appearance of R136a (and the core of NGC 3603)
exhibits broad He\,{\sc ii} $\lambda$4686 emission, which at such 
early phases would not be the case for a lower stellar limit. Other high 
mass
clusters, witnessed at a sufficiently early age, would also betray the 
presence of very massive stars through the presence of He\,{\sc ii} 
emission lines at $\lambda$1640 and $\lambda$4686. Hitherto, such 
clusters may have been mistaken for older clusters exhibiting 
broad helium emission from classical Wolf-Rayet stars.
 
Finally, how might such exceptionally massive stars end their life? This 
question has been addressed from a theoretical perspective (Heger et al. 
2003), suggesting Neutron Star remnants following Type Ib/c core-collapse 
supernovae close to solar metallicities, with weak supernovae and Black 
Hole remnants at LMC compositions. Rapid rotation causes evolution to 
proceed directly to the classical Wolf-Rayet phase, whereas slow rotators, 
such as the NGC 3603 WN6h stars, will likely produce a $\eta$ Car-like 
Luminous Blue Variable phase. Extremely metal-deficient stars exceeding 
$\sim$140 M$_{\odot}$ may end their lives prior to core-collapse (Bond et 
al. 1984). They would undergo an  electron-positron pair-instability 
explosion  during the advanced stages that would trigger the 
complete disruption of the star (Heger \& Woosley 2002).

Until recently, such events have been expected from solely 
metal-free 
(Population III) stars, since models involving the collapse of primordial 
gas clouds suggest preferentially high characteristic masses as large as 
several hundred M$_{\odot}$ (Bromm \& Larson 2004). Langer et al. (2007) 
have demonstrated that local pair-instability supernovae could be produced 
either by slow rotating moderately metal-poor ($\leq$ 1/3 Z$_{\odot}$) 
yellow hypergiants with thick hydrogen-rich envelopes -- resembling SN 
2006gy (Ofek et al. 2007; Smith et al. 2007) -- or rapidly rotating very 
metal-deficient ($\leq 10^{-3}$ Z$_{\odot}$) Wolf-Rayet stars. Therefore 
it is unlikely any of the stars considered here are candidate 
pair-instability supernovae. Nevertheless, the potential for stars 
initially exceeding 140 M$_{\odot}$ within metal-poor galaxies suggests 
that such pair-instability supernovae could occur within the local 
universe, as has recently been claimed for SN 2007bi (Gal-Yam et al. 2009, 
see also Langer 2009).

Finally, close agreement between our spectroscopically derived mass-loss 
rates and theoretical predictions allows us to synthesise the ZAMS 
appearance of very massive stars. We find that the very highest mass 
progenitors will possess an emission-line appearance at the beginning of 
their main-sequence evolution due to their proximity to the Eddington 
limit. As such, spectroscopic dwarf O2--3 stars are not anticipated with 
the very highest masses. Of course the only direct means of establishing 
stellar masses is via close binaries. For the LMC metallicity, Massey et 
al. (2002) have obtained dynamical masses of a few systems, although 
spectroscopic and photometric searches for other high-mass, eclipsing 
binaries in 30 Doradus are in progress through the VLT-FLAMES Tarantula 
Survey (Evans et al. 2010a) and other studies (O. Schnurr et al. in 
prep.). Nevertheless, it is unlikely that any other system within 30 
Doradus, or indeed the entire Local Group of galaxies will compete in mass 
with the brightest components of R136 discussed here.

\section*{Acknowledgements}

We are grateful to C. J. Evans for providing VLT MAD images of R136 and 
photometric zero-points in advance of publication, and E. Antokhina, J. 
Ma\'{i}z Apell\'{a}niz, J. M. Pittard and I. R. Stevens for helpful 
discussions. We thank C. J. Evans and N. R. Walborn and an anonymous 
referee for their critical reading of the manuscript. Based on 
observations made with ESO Telescopes at the Paranal Observatory during 
MAD Science Demonstration runs SD1 and SD2, plus programme ID's 69.D-0284 
(ISAAC), 075.D-0577 (SINFONI), 076-D.0563 (SINFONI). Additional 
observations were taken with the NASA/ESA Hubble Space Telescope, obtained 
from the data archive at the Space Telescope Institute. STScI is operated 
by the association of Universities for Research in Astronomy, Inc. under 
the NASA contract NAS 5-26555. Financial support was provided to O. 
Schnurr and R.J. Parker by the Science and Technology Facilities Council. 
N. Yusof and H.A. Kassim gratefully acknowledge the University of Malaya 
and Ministry of Higher Education, Malaysia for financial support, while R. 
Hirschi acknowledges support from the World Premier International Research 
Center Initiative (WPI Initiative), MEXT, Japan. A visit by N. Yusof to 
Keele University was supported by UNESCO Fellowships Programme in Support 
of Programme Priorities 2008-2009.

\label{lastpage}

\end{document}